\documentclass[a4paper,12pt]{article}
\usepackage[utf8]{inputenc}
\usepackage[english]{babel}
\usepackage{amsmath, amssymb,graphicx,bm}
\usepackage{braket}
\usepackage{caption}
\usepackage{subcaption}
\usepackage{ulem}
\usepackage{multicol}

\pdfoutput=1
\usepackage{jheppub}
\newcommand{\be}{\begin{equation}}
\newcommand{\ee}{\end{equation}}
\newcommand{\bea}{\begin{eqnarray}}
\newcommand{\eea}{\end{eqnarray}}

\newcommand{\cS}{\mathcal{S}}
\newcommand{\cA}{\mathcal{A}}

\newcommand{\fb}{\mathfrak{b}}

\definecolor{darkraspberry}{rgb}{0.53,0.15,0.34}

\definecolor{darkblue}{rgb}{0,0,1}

\definecolor{dgreen}{rgb}{0,0.6,0}

\definecolor{brown}{rgb}{0.59,0.29,0}

\numberwithin{equation}{section}

\begin{document}

\title{Holographic model for light quarks in anisotropic hot dense QGP
  with external magnetic field. }

\author{Irina Ya. Aref'eva$^a$, Alexey Ermakov$^a$, Kristina Rannu$^a$
  and Pavel Slepov$^a$}

\affiliation{$^a$Steklov Mathematical Institute, Russian Academy of
  Sciences,\\ Gubkina str. 8, 119991, Moscow, Russia}

\emailAdd{arefeva@mi-ras.ru}
\emailAdd{ermakov.av15@physics.msu.ru}
\emailAdd{rannu-ka@rudn.ru}
\emailAdd{slepov@mi-ras.ru}

\abstract{We present a five-dimensional twice  anisotropic holographic
  model supported by Einstein-dilaton-three-Maxwell action describing
  light quarks. The first of the Maxwell fields provides finite
  chemical potential. The second Maxwell field serves for anisotropy,
  characterizing spatial anisotropy of the QGP produced in heavy-ion
  collisions (HIC). The third Maxwell field is related to a magnetic
  field that appears in HIC. The dependence of the 5-dim black hole
  solution and confinement/deconfinement phase diagram on this
  magnetic field is considered. The effect of the inverse magnetic
  catalyses for light quarks phase diagram is obtained. Positions
  of critical end points are found. We also study the behavior of the
  conductivity for light quarks in both isotropic and anisotropic
  cases and show that behaviour of the conductivity near critical
  points essentially depend on quark masses, meanwhile at high
  temperature they are similar.}
  


\maketitle

\newpage


\section{Introduction}

Holographic duality provides an effective approach to  study
nonperturbative properties of QCD \cite{Casalderrey-Solana:2011dxg,
  Arefeva:2014kyw, DeWolfe:2013cua}. One of the achievements of the
QCD holographic description is that holographic QCD (HQCD) models can
describe QCD at all energy scales \cite{Gursoy:2008za,
  Gursoy:2010fj}. HQCD reproduces properties of QCD obtained by others
methods, namely perturbative renormalization group  and lattice
approach. However neither of these methods can describe QCD with
non-zero chemical potential and therefore predict the phase diagram
for quark-gluon plasma (QGP). Nowadays  this goal is achieved by
holographic methods only. There are HQCD models that describe QCD with
heavy quarks \cite{Yang:2015aia}, light quarks \cite{Li:2017tdz} as
well as with realistic quarks contents \cite{Pirner:2009gr}. These
models are based on the so-called ``bottom-up'' holographic
approach. Holographic method can also describe anisotropic dense
matter in external magnetic field. \\

The goal of this paper is to study the behaviour of anisotropic  QCD
with light quarks in strong external magnetic field. This paper is the
generalization of our consideration of anisotropic QCD with heavy
quarks in external magnetic field \cite{Arefeva:2020vae} on the one
hand and of our consideration of anisotropic QCD with light quarks at
zero magnetic field \cite{Arefeva:2020byn} to the case of non-zero
external field on the other hand. \\

The  model under consideration \cite{Arefeva:2022bhx} has been built
using a potential reconstruction method
\cite{Casalderrey-Solana:2011dxg, Arefeva:2014kyw, DeWolfe:2013cua, 
  Gursoy:2008za, Gursoy:2010fj, He:2013qq, Yang:2015aia, Li:2017tdz,
  Arefeva:2018hyo, Arefeva:2018cli, Chen:2018msc, Arefeva:2020vhf,
  Ballon-Bayona:2020xls, Arefeva:2020byn, Arefeva:2021kku,
  Slepov:2021gvl, Rannu:2021pcq, Li:2020hau, Arefeva:2021btm,
  Arefeva:2021mag, Arefeva:2020vae, Gursoy:2020kjd, Arefeva:2020bjk,
  Zhou:2021nbp, Dudal:2021jav, Arefeva:2021jpa, Caldeira:2021izy,
  Ali-Akbari:2021ene, Chen:2021gop}. This model is characterized by a
5-dimensional Einstein-dilaton-Maxwell action and a 5-dimensional
metric with a warp factor that tunes the blackening function and,
consequently, the thermodynamics of the model. The phase transition
structure of HQCD strongly depends on the warp factor choice. The
simplest warp factor has an exponential form with a quadratic
polynomial \cite{Andreev:2006nw, Yang:2015aia}. It reproduces the
first-order phase transition from lattice calculations (Columbia plots
\cite{Brown:1990ev, Philipsen:2016hkv}) for the heavy quarks. To
reproduce the first-order phase transition for the light quarks one
has to use rational functions \cite{Gursoy:2010fj, Li:2017tdz}. \\

To describe the QGP produced in heavy ion collisions (HIC) by
holographic methods  one has to make more modifications of the
5-dimensional metric. These modifications are related to two types of
the anisotropy that take  place in HIC: \linebreak i) anisotropy
between the longitudinal and transversal directions and ii) anisotropy
in the transversal plane related to magnetic field produced in
non-central collisions of ions. It was proposed to characterize the
first type of anisotropy by a parameter $\nu$ that is similar to the
Lifshitz parameter of the AdS metric modification. The isotropic model
corresponds to $\nu=1$. For the value of about $\nu = 4.5$ the energy
dependence of the entropy produced in the HIC agrees with the
experimental data \cite{ALICE:2015juo} for the energy dependence of
the total multiplicity of created particles
\cite{Arefeva:2014vjl}. The magnetic field created by non-central ion
collisions \cite{Toneev:2016tgj} produces anisotropy in the plane
transverse to the ion collision axis. Therefore, it is important to
consider these types of anisotropy in holographic models
\cite{Rougemont:2015oea, Li:2016gfn, Gursoy:2016ofp, Dudal:2016joz,
  Gursoy:2017wzz, Gursoy:2018ydr, Bohra:2019ebj, He:2020fdi,
  Rodrigues:2020ndy, Arefeva:2021mag, Arefeva:2020vae}. To construct
anisotropic models, one considers an additional Maxwell field to 
support the anisotropy in metrics \cite{Arefeva:2014vjl,
  Arefeva:2018hyo, Arefeva:2018cli, Arefeva:2020vhf, Arefeva:2020uec,
  Arefeva:2020byn, Arefeva:2021kku, Slepov:2021gvl, Rannu:2021pcq,
  Arefeva:2021btm, Arefeva:2021mag, Arefeva:2020vae, Arefeva:2020bjk,
  Arefeva:2021jpa}. Note that calculations in the twice anisotropic
model for the light quarks are much more complicated than for the
heavy quarks \cite{Arefeva:2021mag, Arefeva:2020vae}, since the warp
factor and the kinetic gauge function $f{_1}$ have more complex
form. For the heavy quark model we have found \cite{Arefeva:2020vae}
the effect of the inverse magnetic catalysis (IMC) -- larger absolute
values of the coupling coefficient $c_B$ lead to decrease of the
transition temperature. In the model for the light quarks considered
in this paper we also find the IMC. This work is a generalization of
results obtained in \cite{Arefeva:2020byn, Arefeva:2020vae} to the
case of twice anisotropic holographic light quark model. The results
obtained in \cite{Arefeva:2020byn} are generalized to the case of
present of an external magnetic field, and the results obtained in
\cite{Arefeva:2020vae} to the case of light quarks. \\

We also study electrical transport properties of QGP and its relation
to the thermal direct photon production in the line of previous
studies of these problems for other holographic models \cite{Wu:2013qja,
  CaronHuot:2006te, Patino:2012py, Finazzo:2013efa, Arciniega:2013dqa,
  Iatrakis:2016ugz, Arefeva:2016rob, Avila:2021rcu}. Photons do not
interact with QGP hadronic matter and therefore provide information
about various characteristics of QGP including phase structure at
different time scales.  It is well known that the direct photons (DP)
emission rate is connected to the conductivity of QGP \cite{Erdmenger,
  Kapusta:2006pm}. In this work we  generalize results of
\cite{Arefeva:2021jpa} concerning the conductivity and photon
production to the light quarks case. We show that conductivity both in
longitudinal and transverse directions essentially depends on
anisotropy parameter $\nu$. The DP flow also depends on temperature
and other thermodynamic properties of QGP. Like it was in the heavy
quarks case, for the light quarks  the higher the magnetic field
or/and chemical potential, the higher the conductivity. We will see
that for $\nu = 1$ the conductivity behavior for the heavy and light
quarks is different at near-critical temperatures, but at high
temperatures they both saturate to constant values. For $\nu = 4.5$
the conductivity behavior for the heavy and light quarks is also
different at near-critical temperatures as well as at high
temperatures. Conductivity for the light quarks model along the
collision direction monotonically decreases with temperature while
transversal components have minimum and increase at large
temperatures. Since we expect isotropisation \cite{Giataganas:2012zy,
  Strickland:2013uga, Arefeva:2021kku}, i.e. $\nu \to 1$ in about
$1-5fm/c\sim 10^{-24}s$, we expect essential change of conductivity at
this time scale. \\

The paper is organized as follows. In Sect.\ref{sec:Model} the
holographic model is presented. In Sect.\ref{Sect:therm} the 
thermodynamics of the model and behaviour of the Wilson loops are
described, the confinement/deconfinement phase diagram is
obtained. Sect.\ref{conduct} describes the derivation of the direct
photons and electrical conductivity properties. In
Sect.\ref{Sect:Conc} we summarize our results and discuss future 
directions of investigations. \\


\section{Model} \label{sec:Model}
In this section we present a model suitable for the light quarks' version
of the confinement/deconfinement phase diagram \cite{Arefeva:2020byn} in
magnetic field similar to how it was done in our work
\cite{Arefeva:2020vae}.

\subsection{Metric and EOM}

Here we take the same action as in our previous work
\cite{Arefeva:2020vae}. In Einstein frame it has the following form 
\begin{gather}
  {\cS} = \int \cfrac{d^5x \, \sqrt{-g}}{16\pi G_5} \left[
    R - \cfrac{f_1(\phi)}{4} \ F_{(1)}^2
    - \cfrac{f_2(\phi)}{4} \ F_{(2)}^2
    - \cfrac{f_B(\phi)}{4} \ F_{(B)}^2
    - \cfrac{\partial_{\mu} \phi \partial^{\mu} \phi}{2}
    - V(\phi) \right]. \label{eq:2.01}
\end{gather}
Here we use the anzats, where the non-zero components of the
electro-magnetic field and the field strengths are
\begin{gather}
  A_{\mu}^{(1)} = A_t (z) \delta_\mu^0, \quad
  F_{y_1 y_2}^{(2)} = q, \quad
  F_{x y_1}^{(B)} = q_B. \label{eq:2.02}
\end{gather}
In action \eqref{eq:2.01} $\phi = \phi(z)$ is the scalar field,
$f_1(\phi)$, $f_2(\phi)$ and $f_B(\phi)$ are the coupling functions
associated with the Maxwell fields $A_{\mu}$, $F_{\mu\nu}^{(2)}$ and
$F_{\mu\nu}^{(B)}$ correspondingly, $q$ and $q_B$ are constants and
$V(\phi)$ is the scalar field potential. Thus \eqref{eq:2.01} is the
extended version of the action used in \cite{Arefeva:2018hyo,
  Arefeva:2020byn}, where we add an external magnetic field
$F_{\mu\nu}^{(B)}$. \\

We also use the same metric anzatz \cite{Arefeva:2020vae}:
\begin{gather}
  ds^2 = \cfrac{L^2 \, \fb(z)}{z^2} \left[
    - \, g(z) dt^2 + dx^2 
    + \left( \cfrac{z}{L} \right)^{2-\frac{2}{\nu}} \!\! dy_1^2
    + e^{c_B z^2} \left( \cfrac{z}{L} \right)^{2-\frac{2}{\nu}} \!\!
    dy_2^2
    + \cfrac{dz^2}{g(z)} \right], \label{eq:2.03} \\
  \fb(z) = e^{2{\cA}(z)}. \label{eq:2.04}
\end{gather}
Let us make some notes on the main parameters of metric
\eqref{eq:2.03}. The difference between ``heavy quarks'' and ``light 
quarks'' cases lies in the form of the warp factor $\fb(z)$. For the heavy
quarks we used $\fb(z) = e^{-\frac{cz^2}{2}}$
\cite{Arefeva:2018hyo}. To get the ``light quarks'' version we follow 
\cite{Li:2017tdz} and assume ${\cA}(z) = - \, a \ln (b z^2 + 1)$. \\

Here parameters $\nu$ and $c_B$ have the same meaning as in
\cite{Arefeva:2020vae}. The primary anisotropy parameter $\nu$
describes non-equivalence between longitudinal and transversal
directions. The isotropic case corresponds to $\nu=1$, while $\nu =
4.5$ reproduces the multiplicity of the charged particles production
\cite{Arefeva:2014vjl}. \\

We refer to $c_B$ as to the magnetic field parameter that describes
non-centrality of the heavy ion collisions. Although explicit relation of
$c_B$ to the magnitude of magnetic field is not established, there are
considerations suggesting that $B^2 = - \, c_B$ \cite{Bohra:2019ebj,
  DHoker:2009ixq}. So we have $c_B < 0$ in all our
calculations. Magnetic field parameter $c_B$ is an implicit function 
of the magnetic charge $q_B$ from the anzats \eqref{eq:2.02}. However,
we will not need the explicit form of this relation in what
follows. We usually set $q=q_B=1$ in numerical calculations. The
metric anzats \eqref{eq:2.03} also contains the blackening function
$g(z)$ which will be determined in the subsequent paragraphs. \\

Applying the stationary action principle, we get the same equations of
motion (EOM) as in \cite{Arefeva:2020vae}:
\begin{gather}
  \begin{split}
    \phi'' &+ \phi' \left( \cfrac{g'}{g} + \cfrac{3 \fb'}{2 \fb} -
      \cfrac{\nu + 2}{\nu z} + c_B z \right)
    + \left( \cfrac{z}{L} \right)^2 \cfrac{\partial f_1}{\partial
      \phi} \ \cfrac{(A_t')^2}{2 \fb g} - \\
    &- \left( \cfrac{L}{z} \right)^{2-\frac{4}{\nu}}
    \cfrac{\partial f_2}{\partial \phi} \ \cfrac{q^2 \ e^{-c_Bz^2}}{2
      \fb g} \
    - \left( \cfrac{z}{L} \right)^{\frac{2}{\nu}} \cfrac{\partial
      f_B}{\partial \phi} \ \cfrac{q_B^2}{2 \fb g} \
    - \left( \cfrac{L}{z} \right)^2 \cfrac{\fb}{g} \ \cfrac{\partial
      V}{\partial \phi} = 0,
  \end{split}\label{eq:2.05} \\
  A_t'' + A_t' \left( \cfrac{\fb'}{2 \fb} + \cfrac{f_1'}{f_1} +
    \cfrac{\nu - 2}{\nu z} + c_B z \right) = 0, \label{eq:2.06} \\
  g'' + g' \left(\cfrac{3 \fb'}{2 \fb} - \cfrac{\nu + 2}{\nu z} + c_B
    z \right)
  - \left( \cfrac{z}{L} \right)^2 \cfrac{f_1 (A_t')^2}{\fb}
  - \left( \cfrac{z}{L} \right)^{\frac{2}{\nu}} \cfrac{q_B^2 \
    f_B}{\fb} = 0, \label{eq:2.07} \\
  \fb'' - \cfrac{3 (\fb')^2}{2 \fb} + \cfrac{2 \fb'}{z}
  - \cfrac{4 \fb}{3 \nu z^2} \left( \cfrac{\nu - 1}{\nu} 
    + \left( 1 - \cfrac{3 \nu}{2} \right) c_B z^2
    - \cfrac{\nu c_B^2 z^4}{2} \right)
  + \cfrac{\fb \, (\phi')^2}{3} = 0, \label{eq:2.08} \\
  2  g' \ \cfrac{\nu - 1}{\nu}
  + 3  g \ \cfrac{\nu - 1}{\nu} \left(
    \cfrac{\fb'}{\fb} - \cfrac{4 \left( \nu + 1 \right)}{3 \nu z}
    + \cfrac{2 c_B z}{3} \right) + \left( \cfrac{L}{z}
  \right)^{1-\frac{4}{\nu}} \cfrac{L \, q^2 \, 
    e^{-c_Bz^2} f_2}{\fb} = 0, \label{eq:2.09} \\
  \begin{split}
    2 g' \left( 1 - \cfrac{1}{\nu} + c_B z^2 \right) 
    &+ 3 g \left[ \Big( 1 - \cfrac{1}{\nu} + c_B z^2 \Big) 
      \left(
        \cfrac{\fb'}{\fb} - \cfrac{4}{3 \nu z} + \cfrac{2 c_B z}{3}
      \right)
      - \cfrac{4 \left( \nu - 1 \right)}{3 \nu z} \right] + \\
    &+ \left( \cfrac{L}{z} \right)^{1-\frac{4}{\nu}}
    \cfrac{L\, q^2 \, e^{-c_Bz^2}  \, f_2}{\fb}
    - \left( \cfrac{z}{L} \right)^{1+\frac{2}{\nu}}
    \cfrac{L \,q_B^2 f_B }{\fb} = 0, 
  \end{split}\label{eq:new} \\
  \begin{split}
    \cfrac{\fb''}{\fb} &+ \cfrac{(\fb')^2}{2 \fb^2}
    + \cfrac{3 \fb'}{\fb} \left( \cfrac{g'}{2 g}
      - \cfrac{\nu + 1}{\nu z}
      + \cfrac{2 c_B z}{3} \right)
    - \cfrac{g'}{3 z g}  \left( 5 + \cfrac{4}{\nu} - 3 c_B z^2
    \right) + \\
    &+ \cfrac{8}{3 z^2} \left( 1 + \cfrac{3}{2 \nu} + \cfrac{1}{2
        \nu^2} \right) 
    - \cfrac{4 c_B}{3} \left( 1 + \cfrac{3}{2 \nu} - \cfrac{c_B
        z^2}{2} \right)
    + \cfrac{g''}{3 g} + \cfrac{2}{3} \left( \cfrac{L}{z} \right)^2
    \cfrac{\fb V}{g} = 0.
  \end{split}\label{eq:2.10}
\end{gather}


Turning off the external magnetic field, i.e. putting $c_B = q_B = f_B
= 0$ into \eqref{eq:2.05}--\eqref{eq:2.10}, we get the EOM from
\cite{Arefeva:2020byn}. Normalizing to the AdS-radius, $L = 1$, we get
the EOM from \cite{Arefeva:2018hyo}. Excluding anisotropy,
i.e. putting $\nu = 1$ and $f_2 = 0$, we get the expressions that
fully coincide with the EOM from \cite{Li:2017tdz, Yang:2015aia}. Thus
\eqref{eq:2.05}--\eqref{eq:2.10} are universal anisotropic EOM,
appropriate both for heavy and light quarks description, that include
solution from \cite{Li:2017tdz, Yang:2015aia} as an isotropic
limit. We also consider the general form of the boundary conditions:
\begin{gather}
  A_t(0) = \mu, \quad A_t(z_h) = 0, \label{eq:2.11} \\
  g(0) = 1, \quad g(z_h) = 0, \label{eq:2.12} \\
  \phi(z_0) = 0, \label{eq:2.13}
\end{gather}
where $z_0 = 0$ corresponds to \cite{Li:2017tdz} and $z_0 = z_h$ to
\cite{Arefeva:2018hyo}. The choice of the boundary condition for the
scalar field was discussed in details in \cite{Arefeva:2020byn}.


\subsection{Solution}

Just as it was in previous cases, to solve EOM
\eqref{eq:2.05}--\eqref{eq:2.10} we need to determine the form of the 
coupling function $f_1$. Let us take the same form as for the light
quarks model \cite{Arefeva:2020byn}. Taking into account the ``light
quarks'' warp factor, we get
\begin{gather}
  f_1 = e^{-cz^2-{\cA}(z)} z^{-2+\frac{2}{\nu}}
  = (1 + b z^2)^a \, e^{- c z^2} z^{-2+\frac{2}{\nu}}. \label{eq:2.14}
\end{gather}
Solving \eqref{eq:2.06} with the coupling function \eqref{eq:2.14} and
boundary conditions \eqref{eq:2.11} gives
\begin{gather}
  A_t = \mu \, \cfrac{e^{\left(2c-c_B\right)z^2/2} -
    e^{\left(2c-c_B\right)z_h^2/2}}{1 -
    e^{\left(2c-c_B\right)z_h^2/2}} \ \xrightarrow[c_B \, \to \, 0]{}
  \ \mu \, \cfrac{e^{cz^2} - e^{cz_h^2}}{1 -
    e^{cz_h^2}}, \label{eq:2.15} \\
  A_t(z) = \mu - \rho \,z^2 + \dots \ \Longrightarrow \
  \rho = - \, \cfrac{\mu  \left(2c - c_B\right)}{2 \left(1 -
    e^{\left(2c-c_B\right)z_h^2/2} \right)}. \label{eq:2.16}
\end{gather}
We take $a = 4.046$, $b = 0.01613$, $c = 0.227$ to make our solution
agree with the results from \cite{Li:2017tdz} in the isotropic case,
the results from \cite{Arefeva:2020byn} in the anisotropic case and
the results from \cite{Arefeva:2020vae} for the heavy quarks version. These
values are due to the mass spectrum of $\rho$ meson with its
excitations and lattice results for the phase transition temperature
\cite{Li:2017tdz}. \\

\begin{figure}[t!]
  \centering
  \includegraphics[scale=0.55]{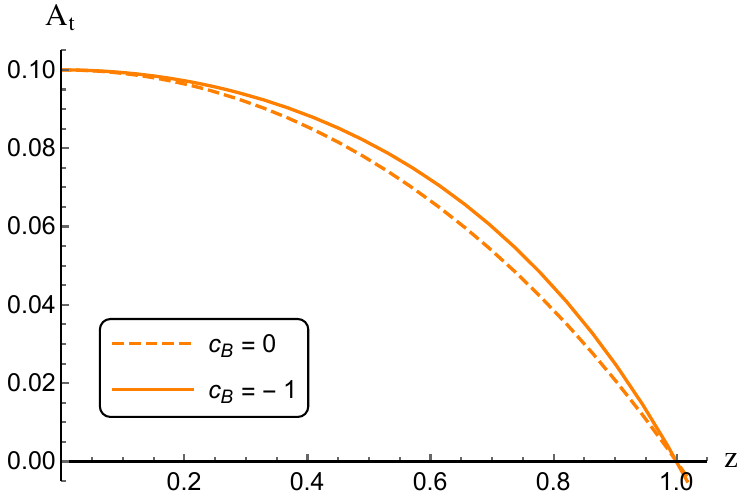} \qquad
  \includegraphics[scale=0.55]{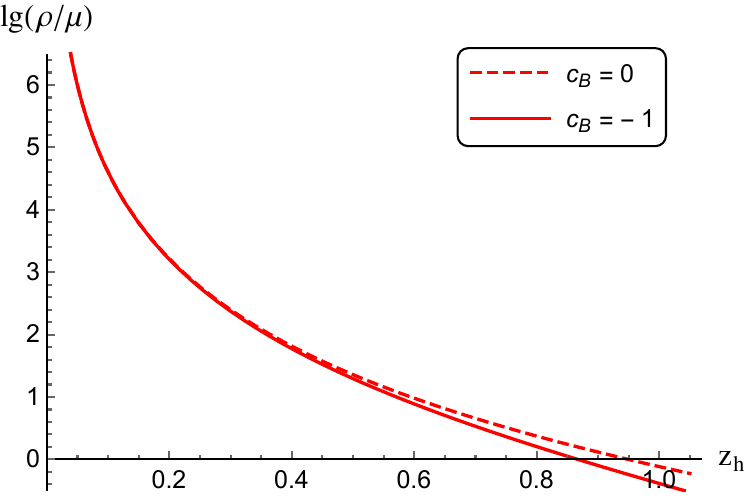} \\
  A \hspace{220pt} B \\
  \caption{Electric potential $A_t(z)$ (A) and density $\rho(z_h)/\mu$
    in logarithmic scale (B) for the ``light quarks'' with (solid lines)
    and without (dashed lines) magnetic field, $c_B = 0, \ - \, 1$
    (B); $\mu = 0.1$, $a = 4.046$, $b = 0.01613$, $c = 0.227$.}
  \label{Fig:Atrho}
\end{figure}

To obtain the blackening function we need to know the kinetic function
$f_B$ describing coupling to the third Maxwell field
$F_{\mu\nu}^{(B)}$. Equation \eqref{eq:2.05} is a consequence of the remaining equations \eqref{eq:2.06}-\eqref{eq:2.10}, which is checked directly. Subtracting \eqref{eq:2.09} from \eqref{eq:new}, we get an expression for the  function $f_B$
\begin{gather}
  f_B = 2 \left( \cfrac{z}{L} \right)^{-2/\nu} \!\!\! \fb g \,
  \cfrac{c_B z}{q_B^2}
  \left( \cfrac{3 \fb'}{2 \fb} - \cfrac{2}{\nu z} + c_B z +
    \cfrac{g'}{g} \right), \label{eq:2.17}
\end{gather}
that allows to transform \eqref{eq:2.07} into
\begin{gather}
  g'' + g' \left(\cfrac{3 \fb'}{2 \fb} - \cfrac{\nu + 2}{\nu z} - c_B
    z \right)
  - 2 g \left(\cfrac{3 \fb'}{2 \fb} - \cfrac{2}{\nu z} + c_B z \right)
  c_B z
  - \left( \cfrac{z}{L} \right)^2 \cfrac{f_1 (A_t')^2}{\fb}  =
  0 \label{eq:2.18} 
\end{gather}
and get the solution
\begin{gather}
  \begin{split}
    &g = e^{c_B z^2} \left[ 1 - \cfrac{I_1(z)}{I_1(z_h)}
      + \cfrac{\mu^2 \, (2 c - c_B)}{L^2 \left( 1 -
          e^{\left(2c-c_B\right)z_h^2/2} \right)^2} \right.  \\
    &\qquad \ \ \left. \times \left\{ \left( e^{\left(2c-c_B\right)z^2/2}
          - e^{\left(2c-c_B\right)z_h^2/2} \right) I_1(z) - (2c - c_B)
        \, I_2(z) \left( 1 - \cfrac{I_1(z)}{I_1(z_h)} \,
          \cfrac{I_2(z_h)}{I_2(z)} \right) \right\} \right], \\
    &I_1(z) = \int_0^{z} \left(1 + b \xi^2 \right)^{3a} e^{- 3 c_B
      \xi^2/2} \, \xi^{1+\frac{2}{\nu}} \, d \xi, \\
    &I_2(z) = \int_0^{z} e^{(2c-c_B)\xi^2/2} \int_0^{\xi} \left(1 + b
      \chi^2 \right)^{3a} e^{- 3 c_B \chi^2/2} \, \chi^{1+\frac{2}{\nu}}
    \, d \chi \, d \xi.
  \end{split}\label{eq:2.19}
\end{gather}

However this solution can't be accepted as the final one and should be
improved. First of all it can't be shown that our previous ``light
quarks'' blackening function \cite{Arefeva:2020byn} serves as zero
magnetic field limit for \eqref{eq:2.19}. To overcome this difficulty
we simplify the expression via partial integration:
\begin{gather}
  g = e^{c_B z^2} \left[ 1 - \cfrac{I_1(z)}{I_1(z_h)}
    + \cfrac{\mu^2 \, (2 c - c_B) \, I_2(z)}{L^2 \left( 1 -
        e^{\left(2c-c_B\right)z_h^2/2} \right)^2} \left( 1 -
      \cfrac{I_1(z)}{I_1(z_h)} \, \cfrac{I_2(z_h)}{I_2(z)} \right)
  \right], \label{eq:2.20}
\end{gather}
where $I_1(z)$ is described by the corresponding expression from
\eqref{eq:2.19}, but $I_2(z)$ was redefined as
\begin{gather}
  I_2(z) = \int_0^{z} \left(1 + b \xi^2 \right)^{3a}
  e^{\left(c-2c_B\right)\xi^2} \, \xi^{1+\frac{2}{\nu}} \, d
  \xi. \label{eq:2.22}
\end{gather} 

Further solving \eqref{eq:2.05}--\eqref{eq:2.10} system with the coupling
functions \eqref{eq:2.14} and \eqref{eq:2.17} and the boundary condition
\eqref{eq:2.13} 
gives

\begin{gather}
  f_2 = 2 \left( \cfrac{z}{L} \right)^{2-4/\nu} \,
  \cfrac{\nu - 1}{q^2 \nu^2 z^2} \,
  \cfrac{e^{c_B z^2}}{(1 + b z^2)^{2a}}
  \left\{ \left( 2 + 2 \nu - \nu c_B z^2 + \cfrac{6 a b \nu z^2}{1 + b
      z^2} \right) g - g' \nu z \right\}, \label{eq:2.23} \\
  \phi = \int_{z_0}^z \sqrt{ \cfrac{4(\nu - 1)}{\nu^2 \xi^2}
    - 2 c_B \left( 3 - \cfrac{2}{\nu} \right) - 2 c_B^2 \xi^2
    + \cfrac{12 a b}{1 + b \xi^2} \left( 1 + 2 \, \cfrac{1 + a b
        \xi^2}{1 + b \xi^2} \right) } \, d \xi, \label{eq:2.24} \\
  \phi' = \sqrt{ \cfrac{4(\nu - 1)}{\nu^2 z^2}
    - 2 c_B \left( 3 - \cfrac{2}{\nu} \right) - 2 c_B^2 z^2
    + \cfrac{12 a b}{1 + b z^2} \left( 1 + 2 \, \cfrac{1 + a b z^2}{1
        + b z^2} \right) }, \label{eq:2.25} \\
  \begin{split}
    &V(z) = - \, \cfrac{(1 + b z^2)^{2a} \ z}{2 L^2} \ \times \\
    & \left[ 2 \left\{ 2 \, \cfrac{ (1 + \nu) (1 + 2 \nu)
            + \bigl( 2 + (2 + 3a ) (3 + 2 \nu) \nu \bigr) b z^2
            + (1 + \nu + 3 a \nu) (1 + 2\nu + 6 a \nu) b^2 z^4}{(1 + b
            z^2)^2 \nu^2 z^2} \right.  \right. \\
      & \left. - \left. c_B \left( 2 + \cfrac{3}{\nu}
            + \cfrac{12 a b z^2}{1 + b z^2} \right) + c_B^2 z^2
        \right\} g \, z
        - \left( 5 + 3 \left( \cfrac{6 a b}{1 + b z^2} - c_B \right)
          z^2 + \cfrac{4}{\nu} \right) g' + g'' z \right].
  \end{split}\label{eq:2.26}
\end{gather}
Note, that we do not fix coupling function for the 3-rd Maxwell $f_B$,
but derive it from the EOM with intent. Fixing, for example, $f_B =
f_1$ makes system \eqref{eq:2.05}--\eqref{eq:2.10} not selfconsistent,
therefore proper solution with the $f_B = f_1$ condition can't be
found. \\


\section{Thermodynamics} \label{Sect:therm}

\subsection{Temperature and entropy}

For the metric \eqref{eq:2.03} and the light quarks warp factor temperature
and entropy can be written as:
\begin{gather}
  T = \cfrac{|g'|}{4 \pi} \, \Bigl|_{z=z_h}, \quad
  s = \left( \cfrac{L}{z_h} \right)^{1+\frac{2}{\nu}}
  \cfrac{e^{c_B z_h^2/2}}{4 \left( 1 + b z_h^2
    \right)^{3a}}. \label{eq:3.03}
\end{gather}

\begin{figure}[t!]
  \centering
  \includegraphics[scale=0.36]{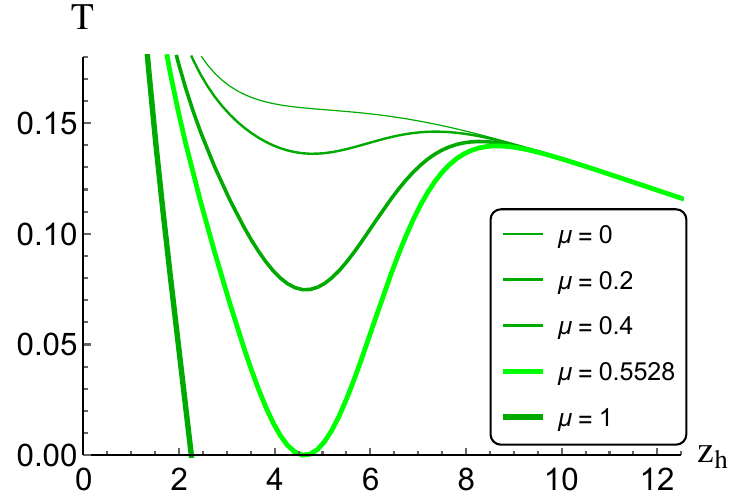} \quad
  \includegraphics[scale=0.36]{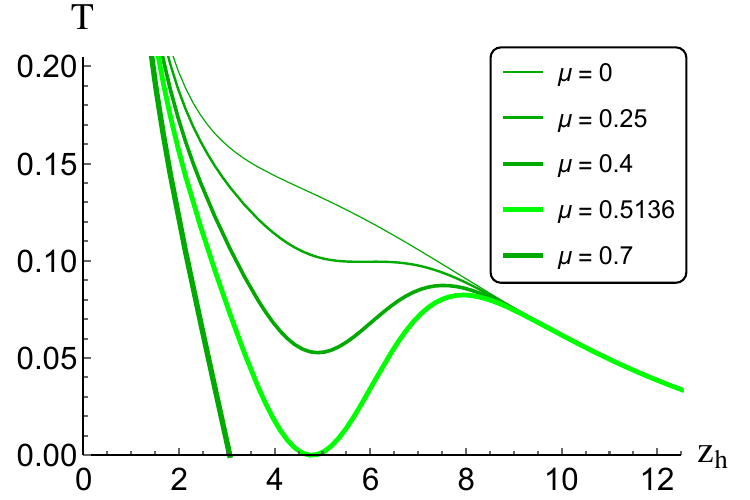} \quad
  \includegraphics[scale=0.36]{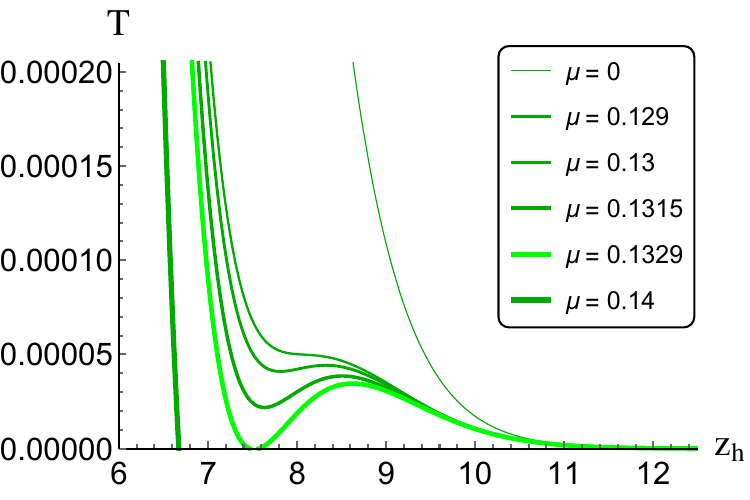} \\
  \includegraphics[scale=0.36]{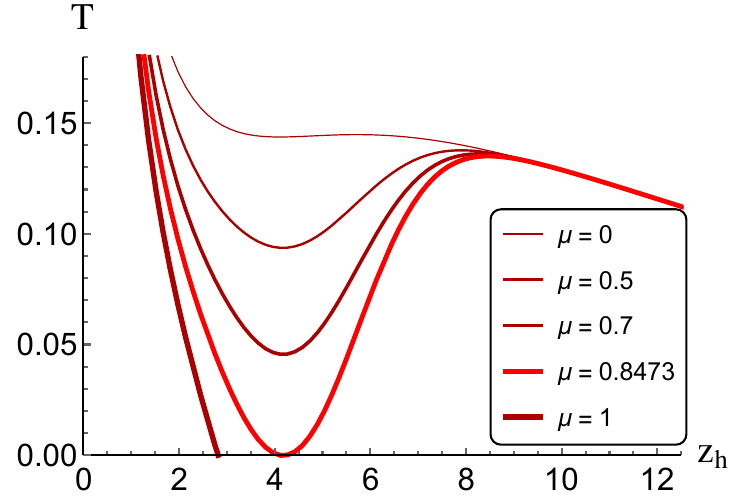} \quad
  \includegraphics[scale=0.36]{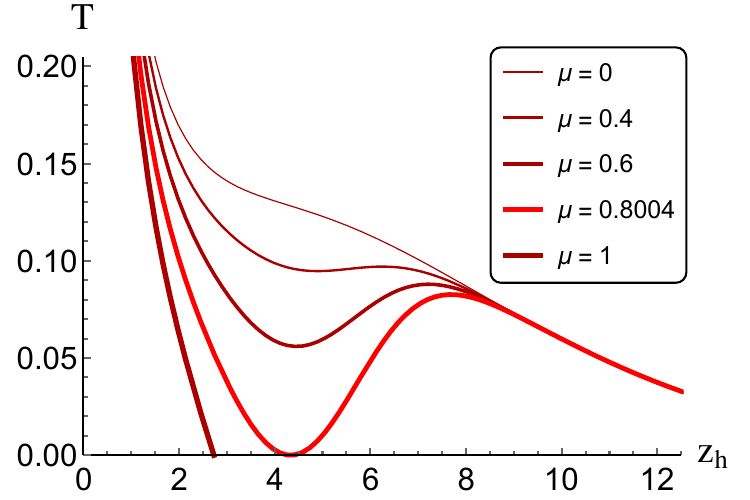} \quad
  \includegraphics[scale=0.36]{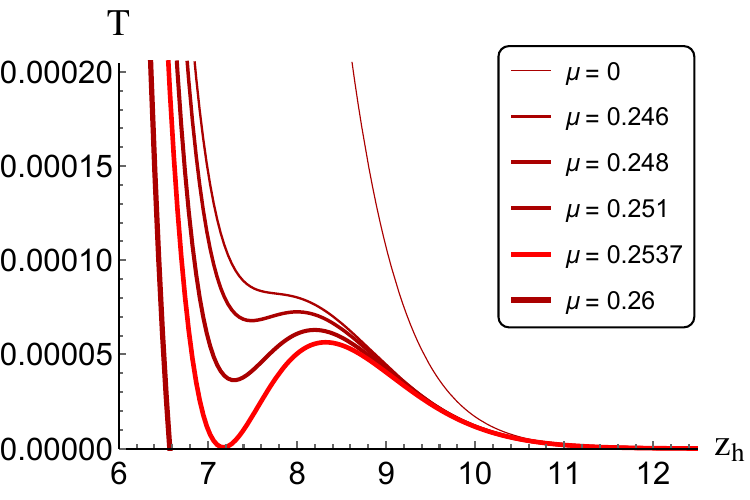} \\
  \includegraphics[scale=0.36]{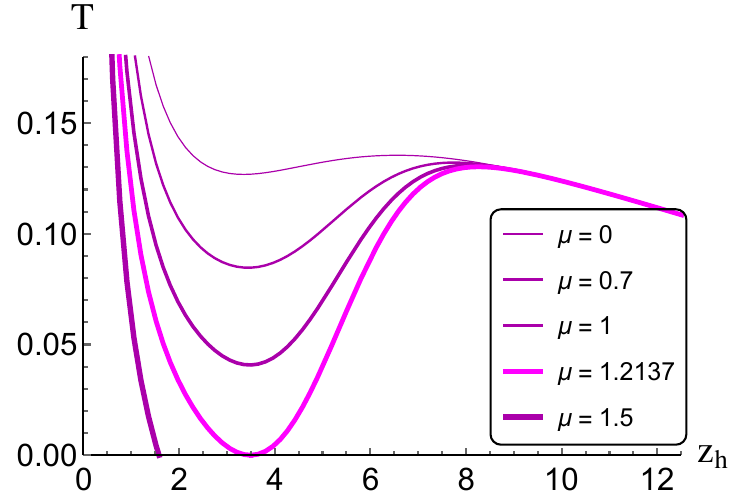} \quad
  \includegraphics[scale=0.36]{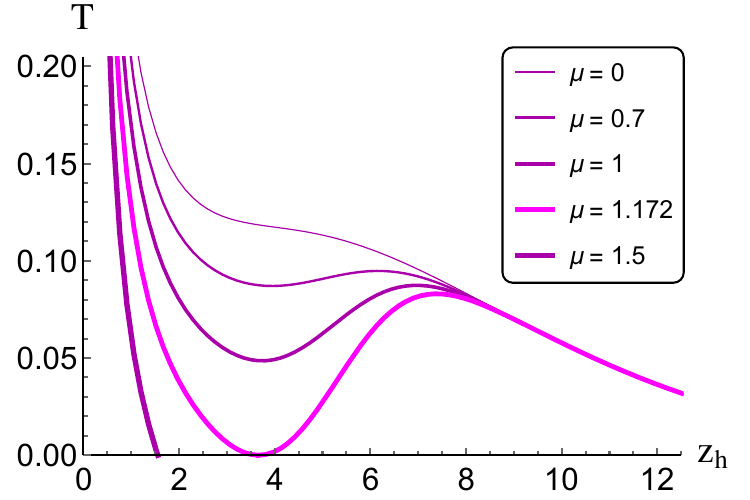} \quad
  \includegraphics[scale=0.36]{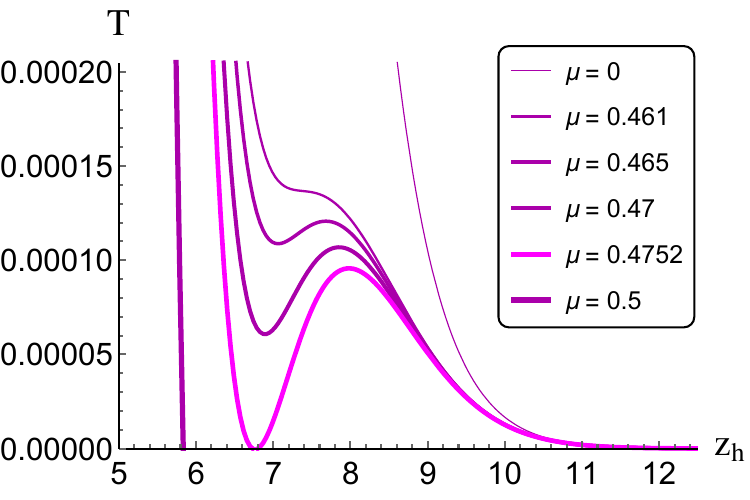} \\
  \includegraphics[scale=0.36]{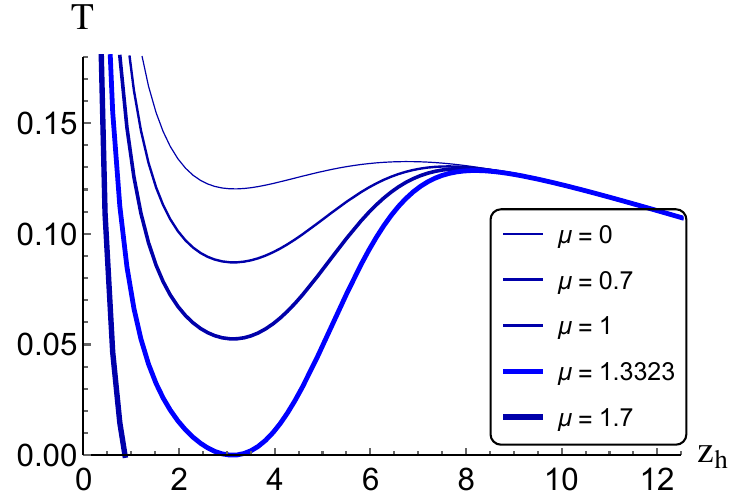} \quad
  \includegraphics[scale=0.36]{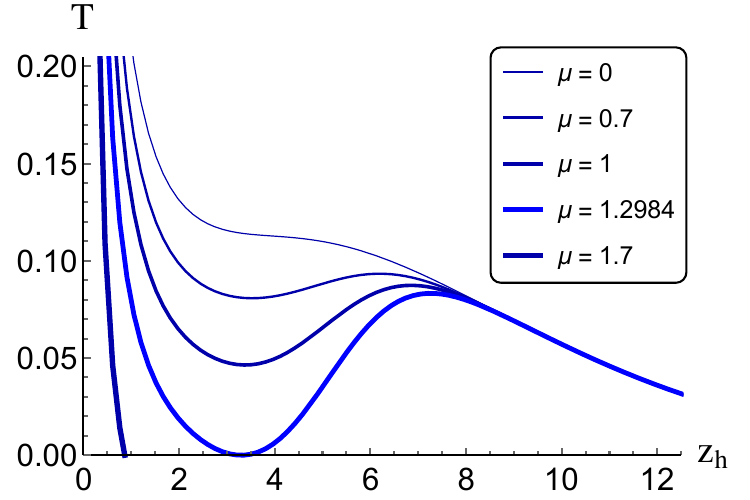} \quad
  \includegraphics[scale=0.36]{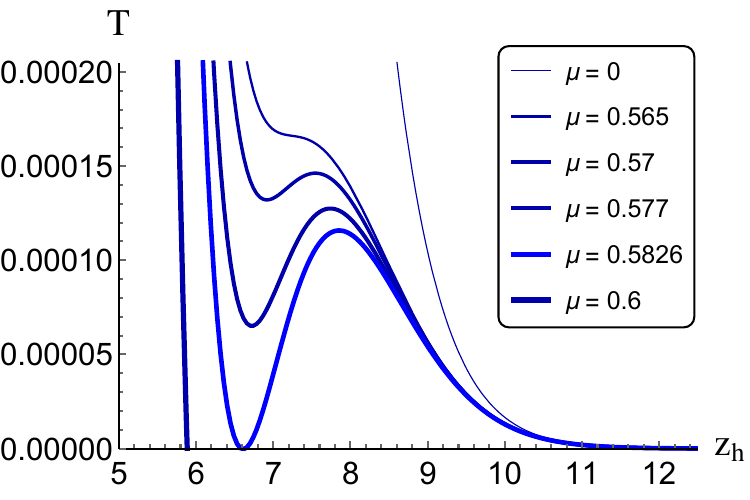} \\
  A \hspace{130pt} B \hspace{130pt} C
  \caption{Temperature as a function of horizon for different $\mu$ in
    isotropic (1-st line, green) and anisotropic cases for $\nu = 1.5$
    (2-nd line, red), $\nu = 3$ (3-rd line, magenta), $\nu = 4.5$
    (4-th line, blue) in magnetic field with $c_B = - \, 0.001$ (A),
    $c_B = - \, 0.01$ (B), $c_B = - \, 0.1$ (C); $a = 4.046$, $b =
    0.01613$, $c = 0.227$. Lighter curves show chemical potentials
    for which second horizon appears and multivalued behavior of
    $T(z_h)$-function actually turns into monotonian.}
  \label{Fig:TScB0}
\end{figure}

On Fig.\ref{Fig:TScB0} temperature as a function of horizon is
presented. The multivalued behavior of $T(z_h)$ preserves starting
from the chemical potential value of the so called critical end point
(CEP) till another specific chemical potential value for which second
horizon appears. For $\nu = 1$ $\mu_{CEP} > 0$, while in cases of
primary anisotropy ($\nu = 1.5, \, 3, \, 4.5$) $\mu_{CEP} = 0$ for
small magnetic fields. Magnetic field's increasing leads to the growth of
$\mu_{CEP}$ values and decrease in characteristic temperatures of the 
multivalued behavior, i.e. the temperatures of the background phase
transition (corresponding to black hole -- black hole collapse in
AdS$_5$). Thus in the magnetic field strong enough background phase
transition should occur at high chemical potentials and negligible
temperatures.


\subsection{Background phase transition}

\begin{figure}[t!]
  \centering
  \includegraphics[scale=0.4]{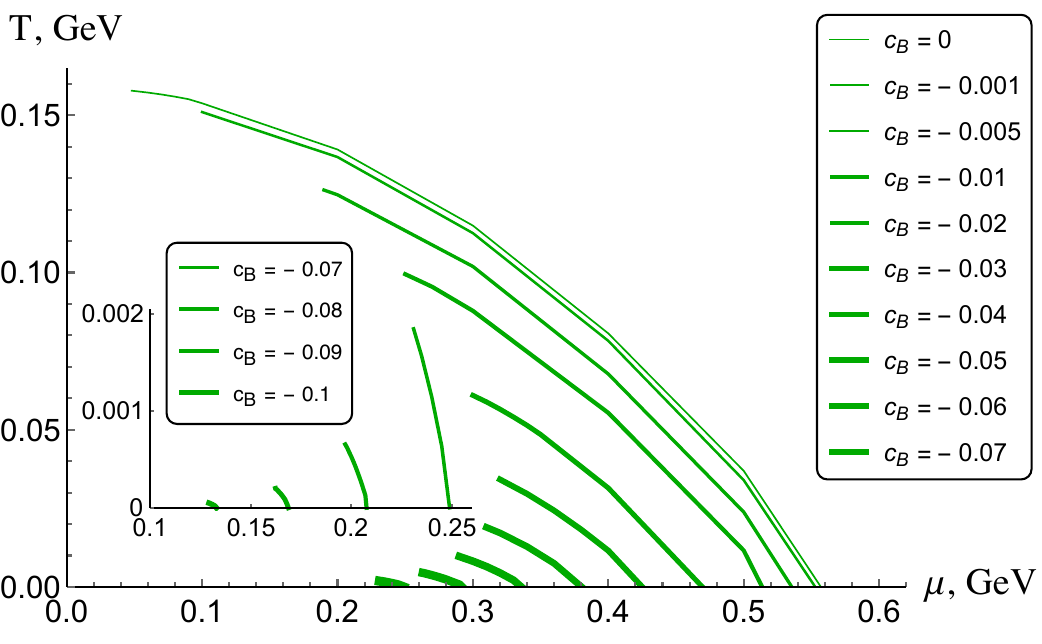} \quad
  \includegraphics[scale=0.4]{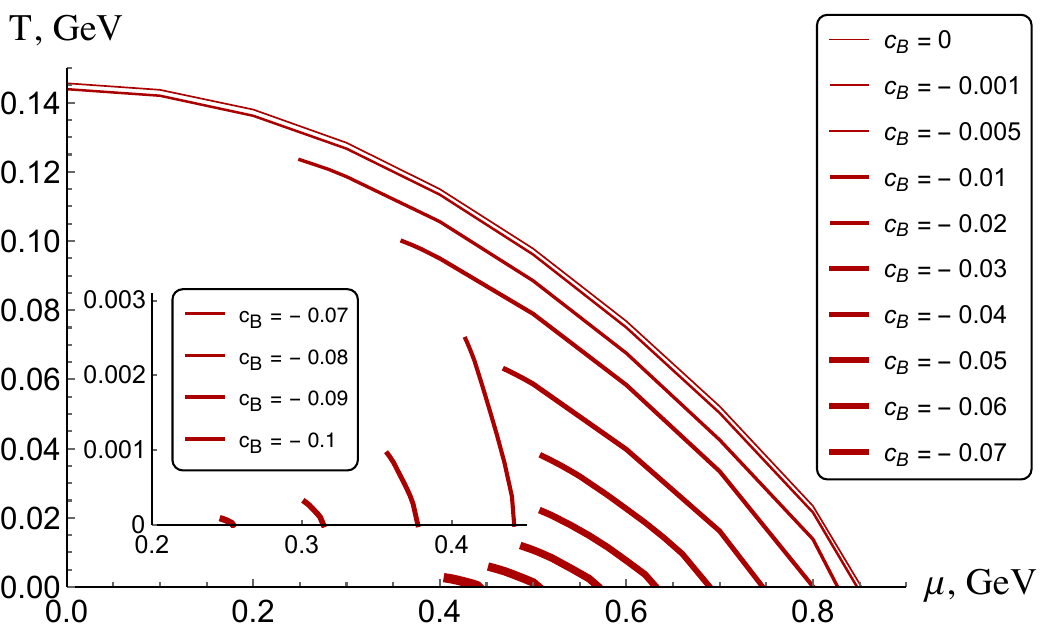} \\
  A \hspace{220pt} B \\
  \includegraphics[scale=0.4]{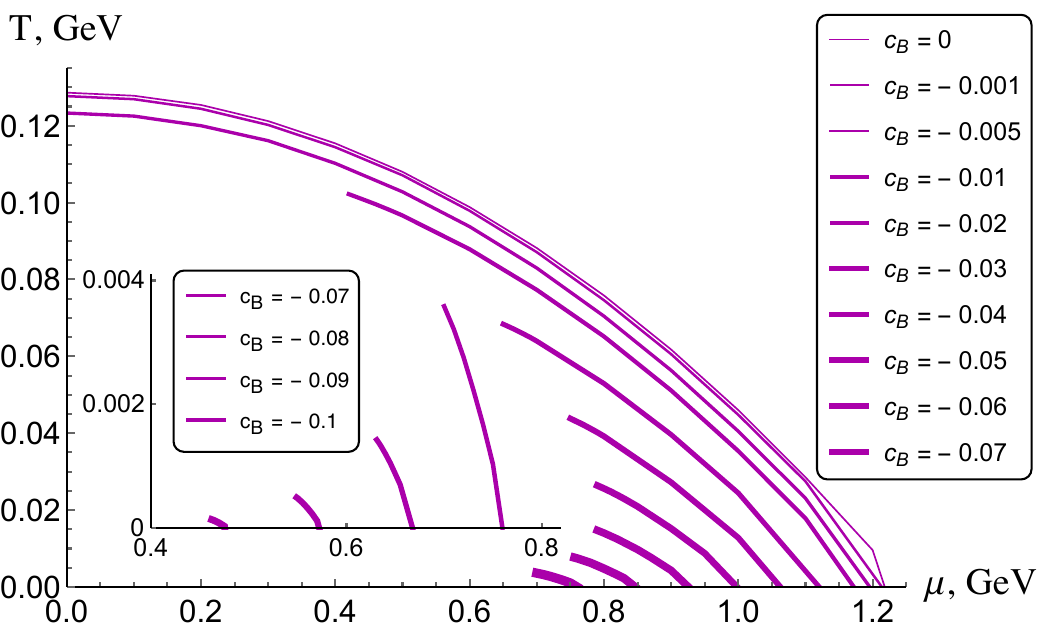} \quad
  \includegraphics[scale=0.4]{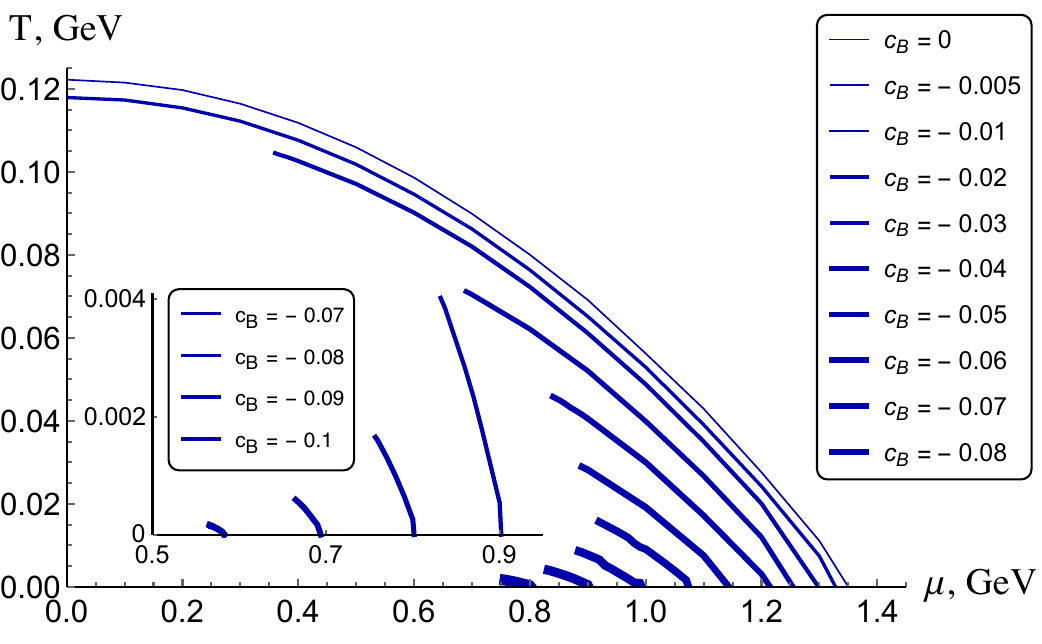} \\
  C \hspace{220pt} D \\ \ \\
  \includegraphics[scale=0.5]{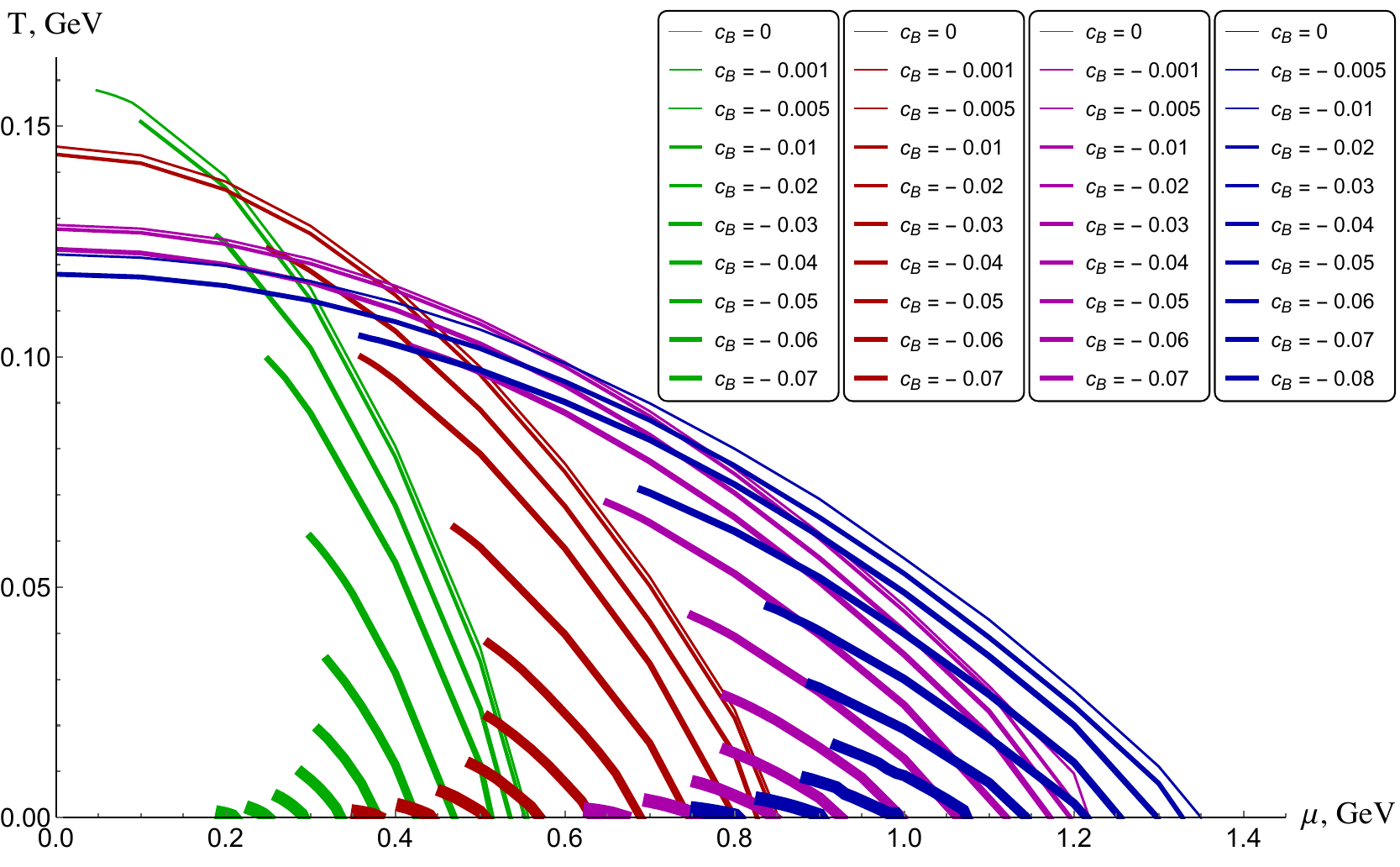} \\
  E 
  \caption{Background phase transition diagram $T(\mu)$ in isotropic
    (A) and anisotropic cases for $\nu = 1.5$ (B), $\nu = 3$ (C), $\nu
    = 4.5$ (D) and all combined (E); $a = 4.046$, $b = 0.01613$, $c =
    0.227$.}
  \label{Fig:Tmu}
\end{figure}

\begin{figure}[b!]
  \centering
  \includegraphics[scale=0.38]{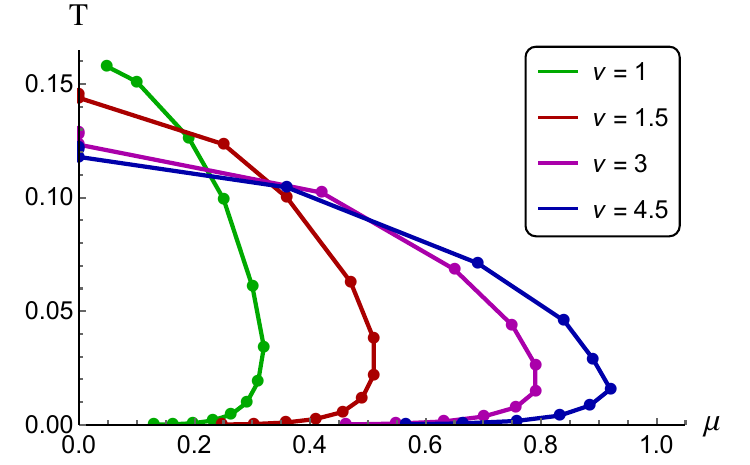} \!\!\!
  \includegraphics[scale=0.38]{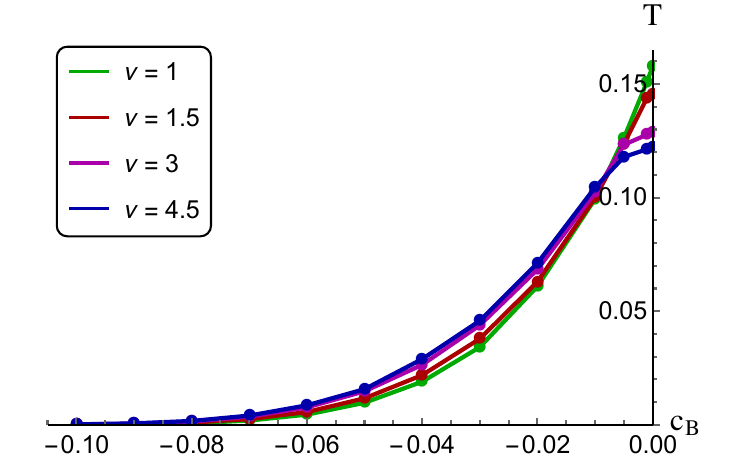} \!\!\!
  \includegraphics[scale=0.38]{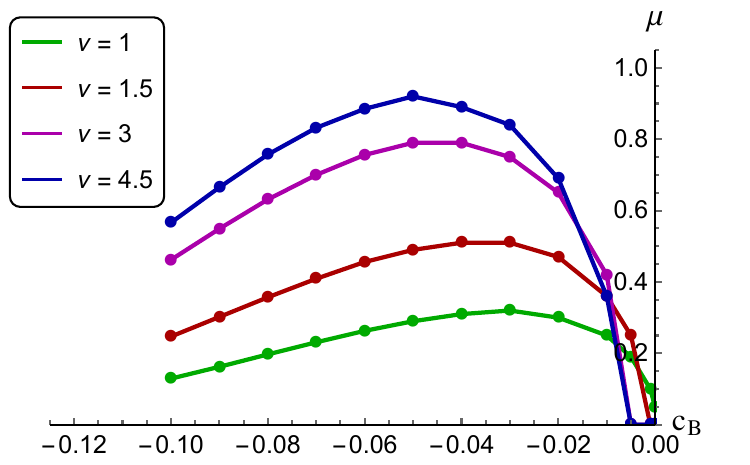} \\
  A \hspace{130pt} B \hspace{130pt} C \\
  \hskip-3em
  \includegraphics[scale=0.3]{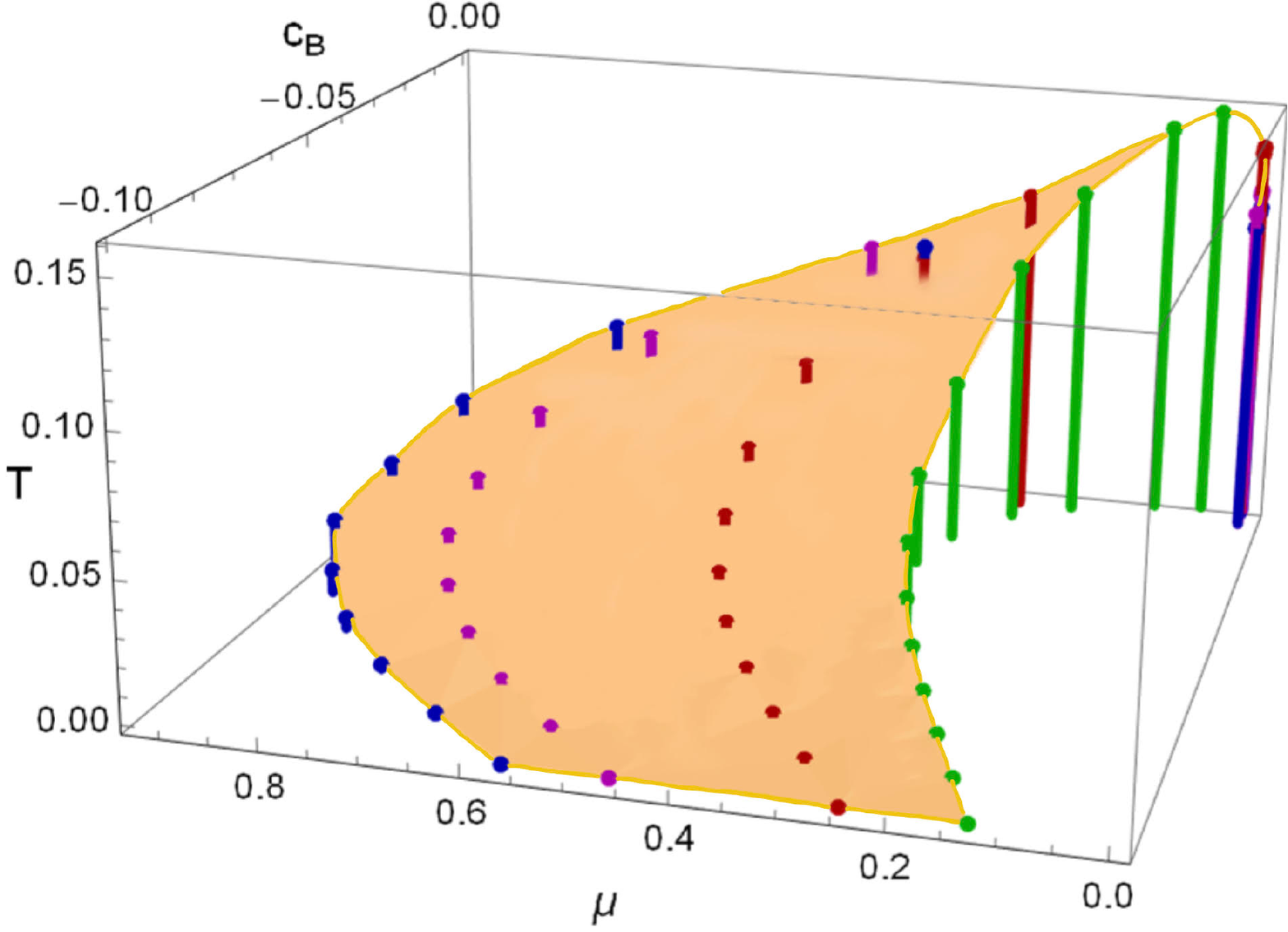}
  D
  \caption{CEP positions on the background phase transition diagram
    $T(\mu)$ in isotropic and anisotropic cases for $\nu = 1, \, 1.5,
    \, 3, \ 4.5$ on planes $\mu-T$ (A), $c_B-T$ (B), $c_B-\mu$ (C) and
    3D plot $T(c_B,\mu)$ (D); $a = 4.046$, $b = 0.01613$, $c =
    0.227$.}
  \label{Fig:CEP}
\end{figure}

This is exactly the picture we observe on phase diagrams
Fig.\ref{Fig:Tmu}, obtained via free energy consideration
\begin{gather}
  F = - \int s \, d T = \int_{z_h}^{\infty} s \, T'
  dz, \label{eq:3.05}
\end{gather}
where we normalize the free energy to vanish at $z_h \to \infty$.
The background phase transition curves become shorter with the
magnetic field increasing (larger $c_B$ absolute values). Note, that
two opposite tendencies can be seen: growth of the $\mu_{CEP}$ and
decrease of the maximum chemical potential value. For $c_B \approx (-
\, 0.05; - \, 0.03)$ depending on $\nu$ these tendencies come into
conflict, and CEP curve  reverses it's movement back towards lower
values of the chemical potential (Fig.\ref{Fig:CEP}A). \\

Primary anisotropy affects the background phase transition for the light
quarks QGP in much the same way as it did for the heavy quarks
\cite{Arefeva:2020vae}: lowers transition temperatures and enlarges
chemical potentials avaliable. But it almost doesn't influence the
limit $c_B$ values at which phase transition still exists. Phase
transition temperatures reach negligible values at $c_B \approx - \,
0.1$ almost regardless of primary anisotropy $\nu$
(Fig.\ref{Fig:Tmu}A-D, Fig.\ref{Fig:CEP}B). It's effect on $c_B$ for
which $\mu_{CEP}$ becomes non-zero is more notable: $c_B < - \, 0.001$
for $\nu = 1.5$ and $\nu < - \, 0.005$ for $\nu = 4.5$
(Fig.\ref{Fig:CEP}A,C). \\

We present CEP for $\nu = 1, \, 1.5, \, 3, \, 4.5$ in a 3-dimensional
space of parameters ($\mu, c_B, T$) on Fig.\ref{Fig:CEP}D. Thus,
figures A, B and C are the projections of a complete picture D on the
corresponding planes. Also, one can see the smooth peach coloured
surface of the first order phase transitions interpolating for
intermediate values of anisotropy. The orange curve indicates
boundaries of possible phase transition locations in this space of
parameters.



\subsection{Temporal Wilson loops}

The expectation value of the temporal Wilson loop calculation does not
differ from the heavy quarks case \cite{Arefeva:2020vae} untill we
substitute the specific light quarks warp factor into the general
expressions for the dynamical wall (DW) equations:
\begin{gather}
  \begin{split}
    DW_x: &\quad - \, \cfrac{4 a b z}{1 + b z^2} 
    + \sqrt{\cfrac23} \ \phi'(z) + \cfrac{g'}{2 g} 
    - \cfrac{2}{z} \ \Big|_{z = z_{DWx}} \hspace{-15pt} = 0, \\
    DW_{y_1}: &\quad - \, \cfrac{4 a b z}{1 + b z^2}
    + \sqrt{\cfrac23} \ \phi'(z) + \cfrac{g'}{2 g}
    - \cfrac{\nu + 1}{\nu z} \ \Big|_{z = z_{DWy_1}} \hspace{-15pt} =
    0, \\
    DW_{y_2}: &\quad - \, \cfrac{4 a b z}{1 + b z^2} 
    + \sqrt{\cfrac23} \ \phi'(z) + \cfrac{g'}{2 g} 
    - \cfrac{\nu + 1}{\nu z} + c_B z \ \Big|_{z = z_{DWy_2}}
    \hspace{-15pt} = 0.
  \end{split}\label{eq:3.06}
\end{gather}

\begin{figure}[b!]
  \centering
  \includegraphics[scale=0.4]{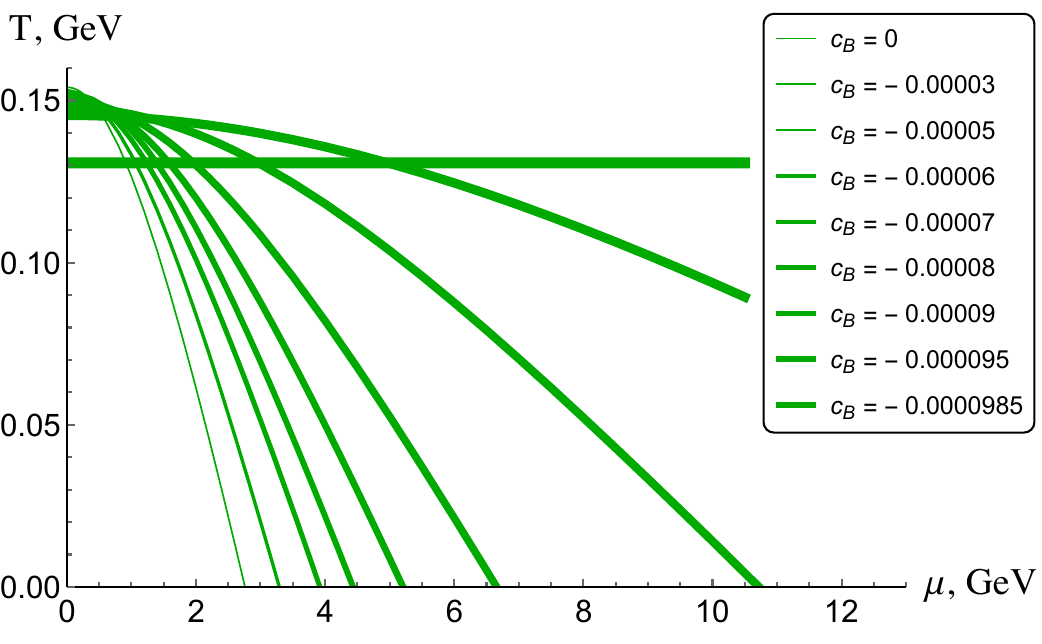} \quad
  \includegraphics[scale=0.4]{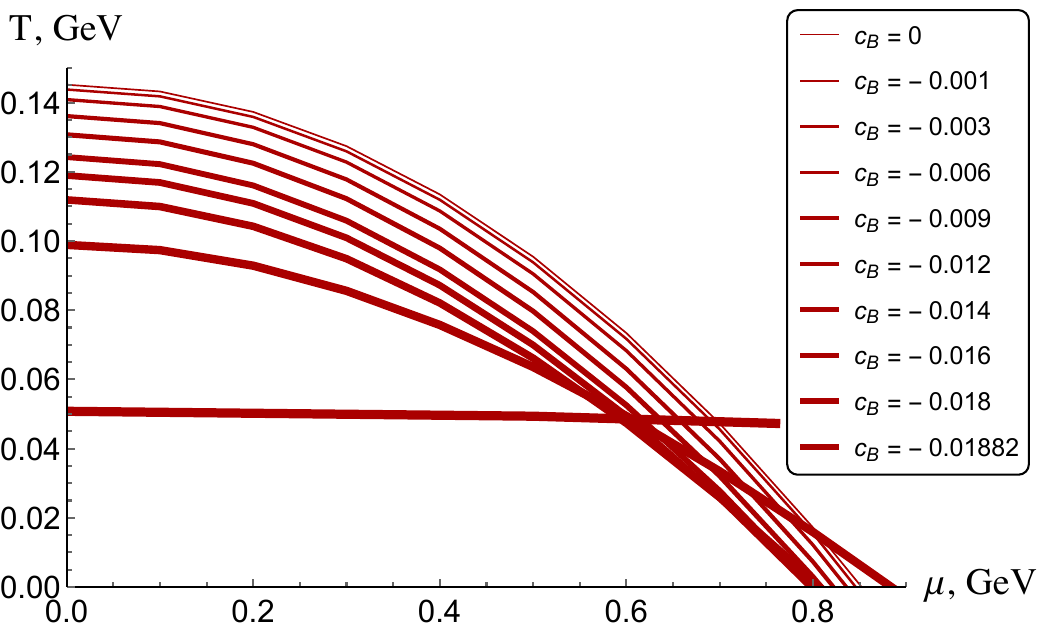} \\
  A \hspace{220pt} B \\
  \includegraphics[scale=0.4]{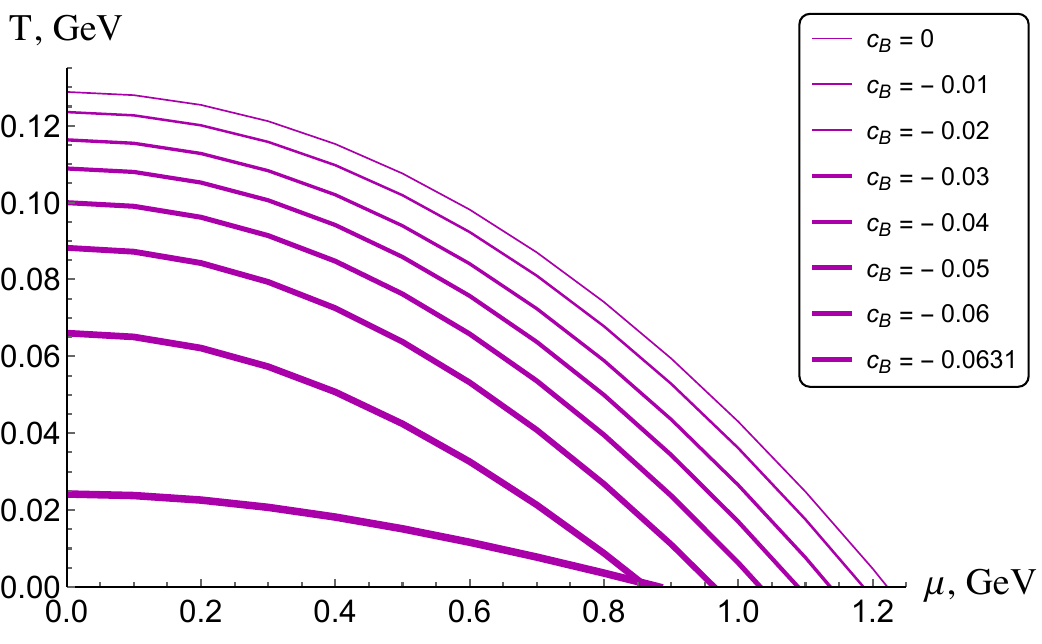} \quad
  \includegraphics[scale=0.4]{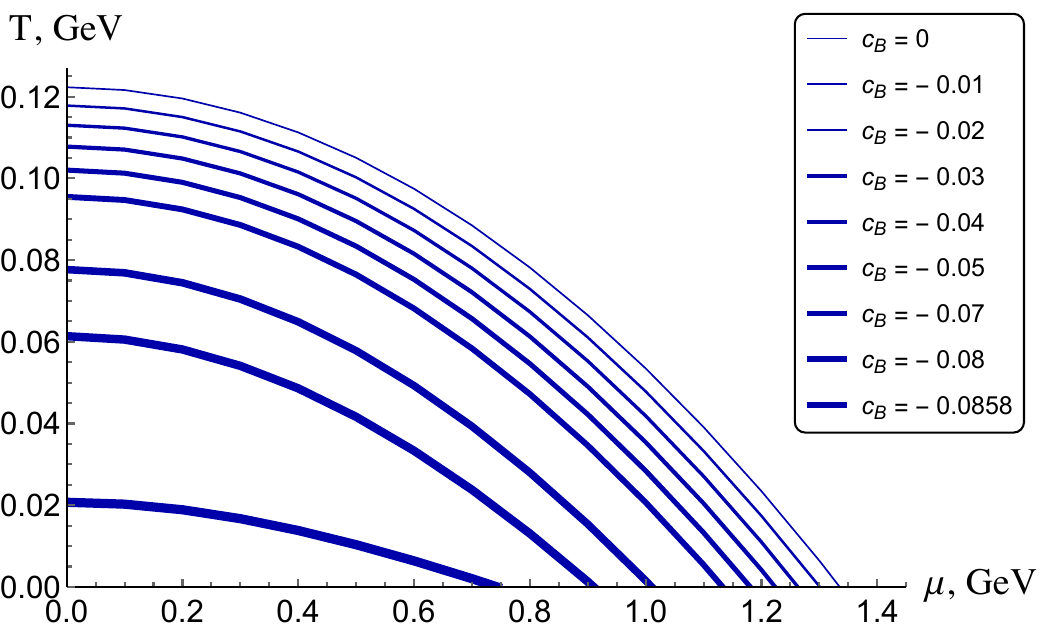} \\
  C \hspace{220pt} D
  \caption{Wilson loop phase transition curves $T(\mu)$ in isotropic
    (A) and anisotropic cases for $\nu = 1.5$ (B), $\nu = 3$ (C) and
    $\nu = 4.5$ (D); $a = 4.046$, $b = 0.01613$, $c = 0.227$.}
  \label{Fig:WL}
\end{figure}

Primary anisotropy has significant influence on the Wilson loop
transformations in the magnetic field (Fig.\ref{Fig:WL}). For $\nu = 1$
the Wilson loop phase transition curve becomes less inclined and it's temperature
decreases, while maximum chemical potential grows. At $c_B = - \,
0.0000985$ the curve becomes horizontal and for $c_B < - \, 0.0000985$
it just disappears (Fig.\ref{Fig:WL}A). For $\nu = 4.5$ larger
absolute $c_B$ also leads to lesser temperatures, but chemical
potential decreases as well, so the Wilson loop curve shrinks clinging
to the abscissa axis $\mu$ at the end. The curve disappears for $c_B <
- \, 0.0858$ (Fig.\ref{Fig:WL}D). For intermediate values of primary
anisotropy $\nu = 1.5$ (Fig.\ref{Fig:WL}B) and $\nu = 3$
(Fig.\ref{Fig:WL}C) a mixture of both these tendences can be
seen. But inverse magnetic catalysis takes place for any primary
anisotropy. \\

\begin{figure}[t!]
  \centering
  \includegraphics[scale=0.38]{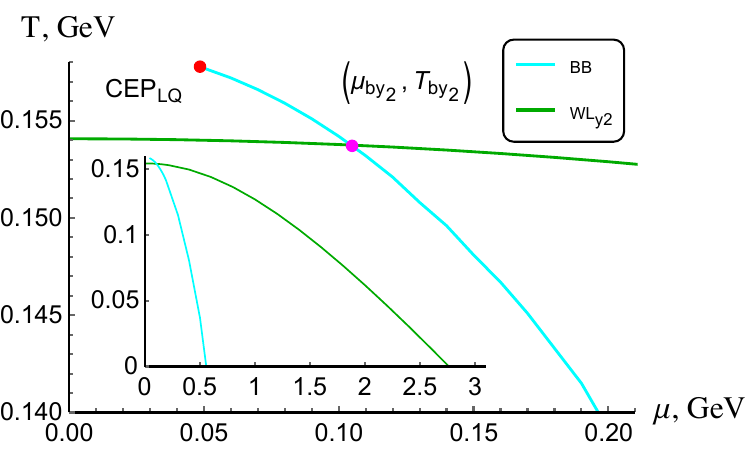} \
  \includegraphics[scale=0.38]{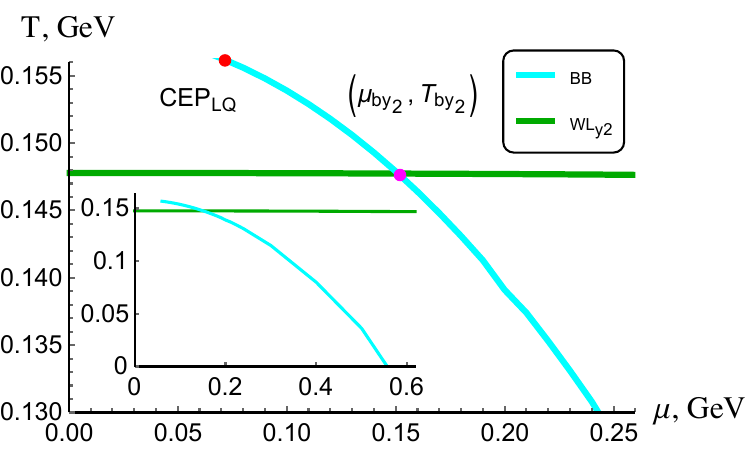} \
  \includegraphics[scale=0.38]{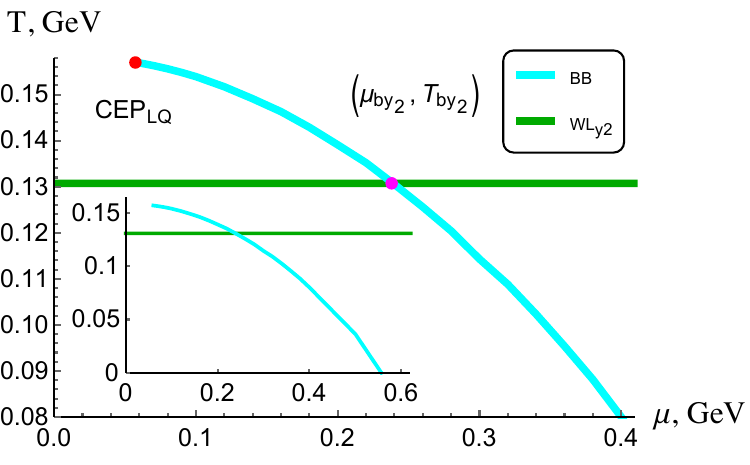} \\
  \includegraphics[scale=0.38]{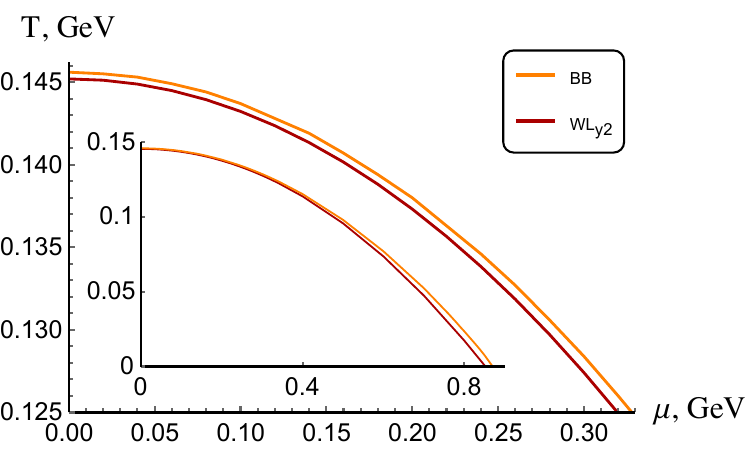} \
  \includegraphics[scale=0.38]{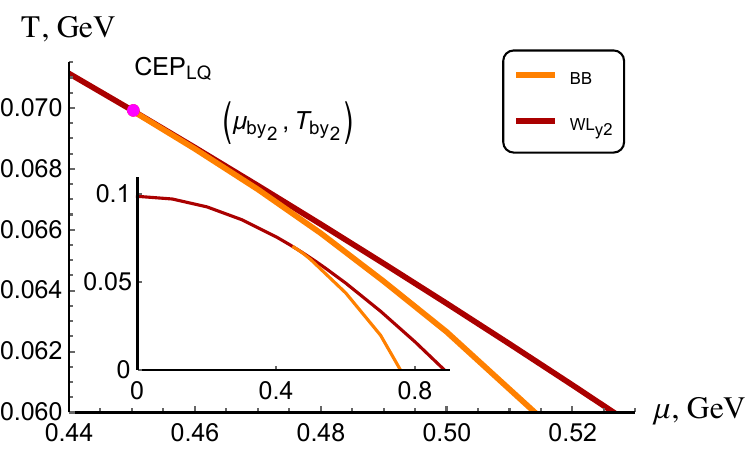} \
  \includegraphics[scale=0.38]{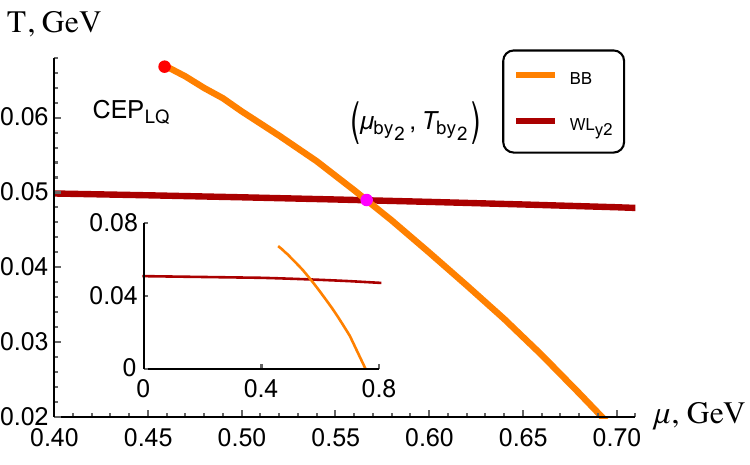} \\
  \includegraphics[scale=0.38]{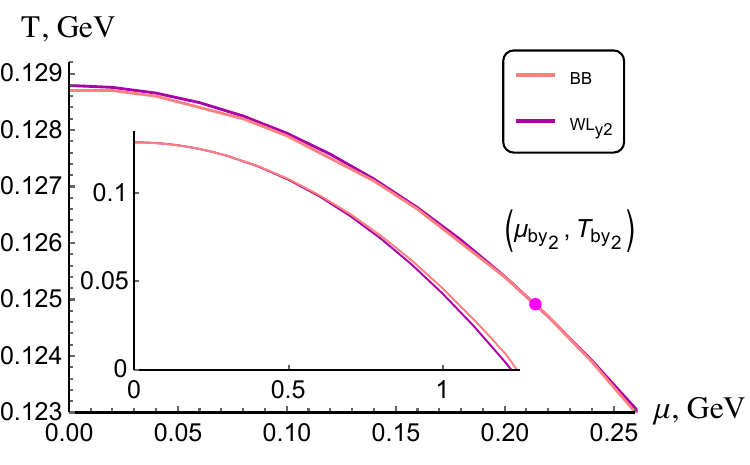} \
  \includegraphics[scale=0.38]{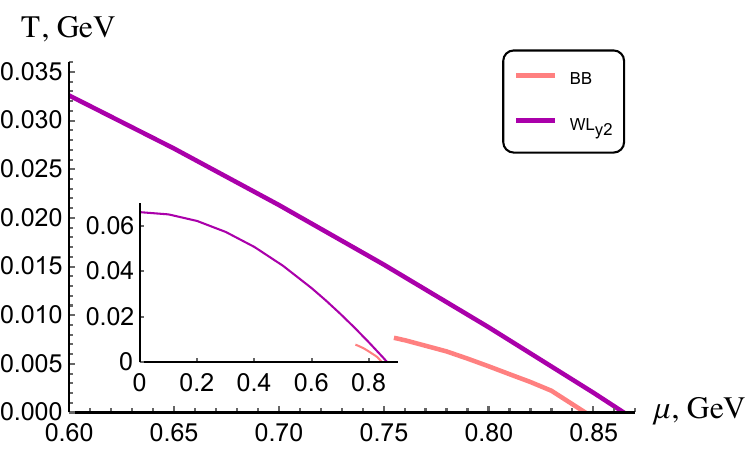} \
  \includegraphics[scale=0.38]{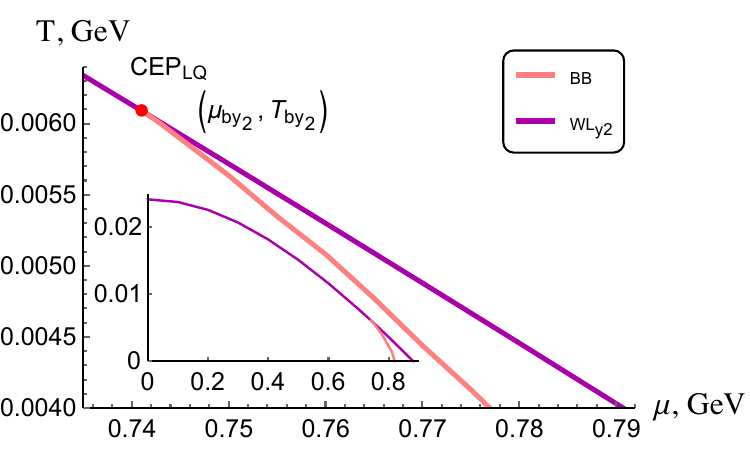} \\
  \includegraphics[scale=0.38]{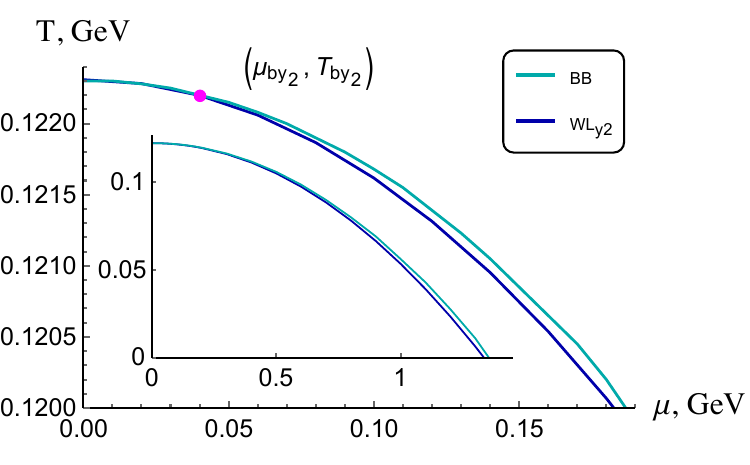} \
  \includegraphics[scale=0.38]{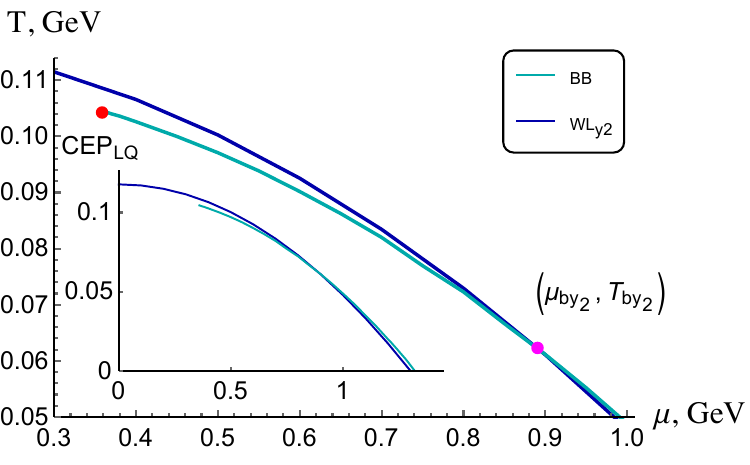} \
  \includegraphics[scale=0.38]{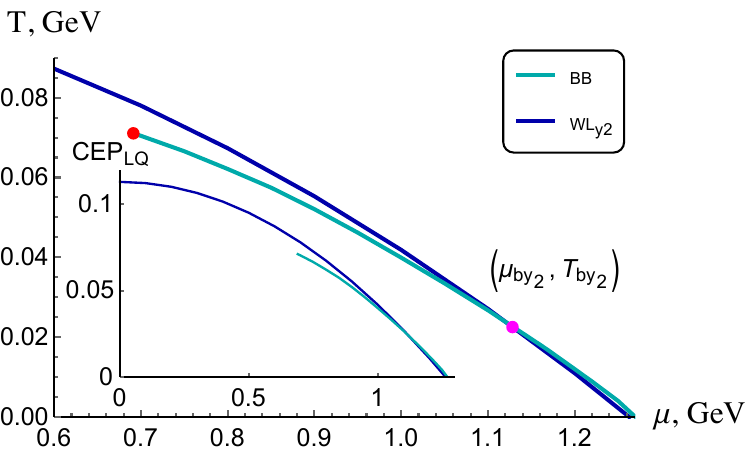} \\
  \includegraphics[scale=0.38]{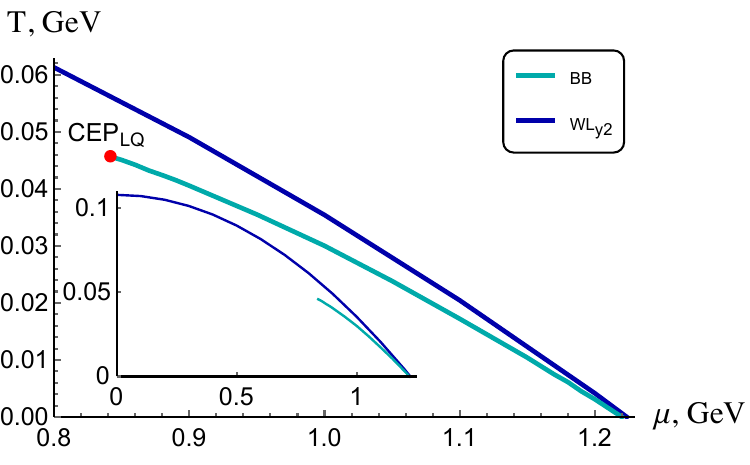} \
  \includegraphics[scale=0.38]{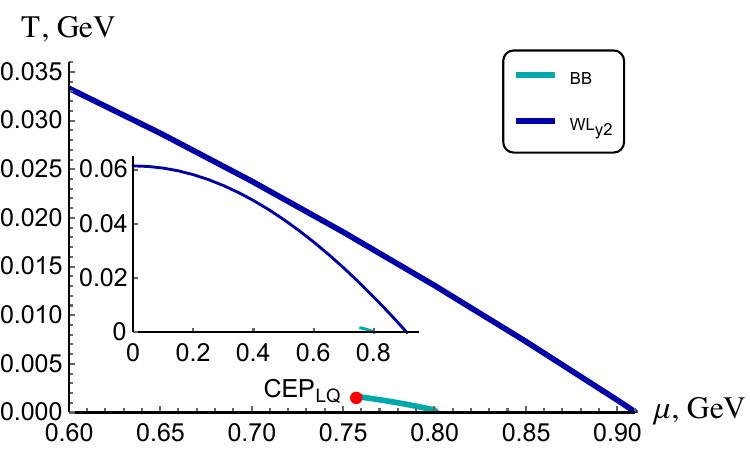} \
  \includegraphics[scale=0.38]{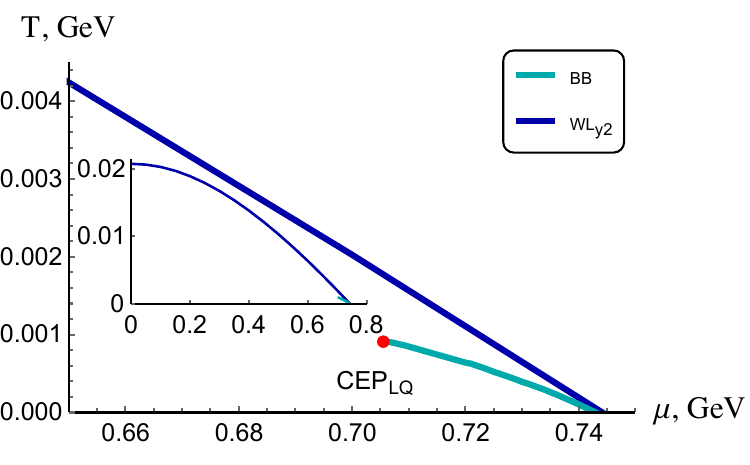} \\
  A \hspace{130pt} B \hspace{130pt} C
  \caption{Phase diagram $(\mu,T)$ as a combination of the 1-st order
    phase transition (BB) and a crossover (WL) in isotropic (1-st
    line, green) with $c_B = 0$ (A), $c_B = - \, 0.00009$ (B), $c_B =
    - \, 0.0000985$ (C), and anisotropic cases for $\nu = 1.5$ (2-nd
    line, red) with $c_B = 0$ (A), $c_B = - \, 0.018$ (B), $c_B = - \,
    0.01882$ (C), $\nu = 3$ (3-rd line, magenta) with $c_B = 0$ (A),
    $c_B = - \, 0.06$ (B), $c_B = - \, 0.0631$ (C), $\nu = 4.5$ (4-th
    and 5-th line, blue), with $c_B = 0$ (A), $c_B = - \, 0.01$ (B),
    $c_B = - \, 0.02$, (C), $c_B = - \, 0.03$ (A), $c_B = - \, 0.08$
    (B), $c_B = - \, 0.0858$ (C); $a = 4.046$, $b = 0.01613$, $c =
    0.227$.}
  \label{Fig:WL-BB}
\end{figure}

Let us now consider the combination of the 1-st order phase transition
and a crossover for the primary anisotropy and magnetic field fixed (while
both transition curves do exist). On Fig.\ref{Fig:WL-BB} regions of
their intersection or approach are shown, and the general view of the
phase diagrams can be found on the inserted mini-plots. Points of
intersections, where the confinement/deconfinement main role passes
from the crossover to the 1-st order phase transitions (or vice
versa), are marked as $(\mu_{by_2}, T_{by_2})$
(Tab.\ref{Tab:BB-WL}). Critical end points, where the 1-st order phase
transition begins, are marked as CEP$_{LQ}$. \\

For the primary isotropic case $\nu = 1$ a crossover disappears very
soon, so it's position relative to the 1-st order phase transition
does not have time to change much. The intersection point just shifts
to the larger $\mu$ and the lesser $T$. \\

For $\nu = 1.5$ both curves are very close to each other and look
cocentric. They do not intersect, so the crossover defines the phase
transition for all chemical potential values. Magnetic field makes
these curves closer to each other and hardly distinguishable, but the
intersection appears at about $c_B = - \, 0.018$ and coincides with
the CEP$_{LQ}$. At this moment phase transition main role clearly
passes from the crossover to the 1-st order phase transition though for
rather a short interval. Before it's disappearance at $c_B = - \,
0.01882$ the crossover line shifts down to lesser temperatures and
becomes almost horizontal like it was for $\nu = 1$. \\

For $\nu = 3$ the 1-st order phase transition and a crossover can't be
distiguished properly. Magnetic field shifts the 1-st order phase
transition curve under the crossover, so that they do not
intersect. Under these circumstances there is no smooth transfer
between the crossover and the 1-st order phase transition, but a jump
to lower temperature -- temperature of CEP$_{LQ}$. Near the limit $c_B
= - \, 0.0631$ the crossover catches up with the 1-st order phase
transition and a smooth transfer appears again just before the
crossover vanishes. \\

\newpage

\

\newpage

For the maximal primary anisotropy $\nu = 4.5$ the 1-st order phase
transition curve constantly shifts under the crossover curve, so that
the intersection point moves to larger chemical potentials and lower
temperatures. On the other hand CEP$_{LQ}$ also moves, causing a jump
transfer from the crossover to the 1-st order phase transition. The
reverse transition is smooth. For $c_B = - \, 0.03$ the 1-st order
phase transition is completely under the crossover so that the reverse
transition does not occur, and further for $c_B < - \, 0.03$ it occurs
as a jump to higher temperatures. The 1-st order phase transition
exists in a very narrow interval. So the confinement/deconfinement
transition should be defined by a crossover with constantly decreasing
temperature except a small local region of almost zero $T$. Near the
limit value $c_B = - \, 0.0858$ the crossover catches up with the 1-st
order phase transition like it was for $\nu = 3$. Soon after this
moment the crossover vanishes, but the 1-st order phase transition
doesn't survive for long after that. \\

\begin{table}
  \centering 
  \begin{tabular}{|c|c|c|}
    \hline
    \ & $c_B$ & $(\mu_{by_2}, T_{by_2})$ \\
    \hline
    \ & $0$ & $(0.1049, 0.1531)$ \\
    $\nu = 1$ & $- \, 0.00009$ & $(0.1405, 0.1479)$ \\
    \ & $- \, 0.0000985$ & $(0.2308, 0.1311)$ \\
    \hline
    \ & $0$ & no intersection \\
    $\nu = 1.5$ & $- \, 0.018$ & $(0.4501, 0.0699)$ \\
    \ & $- \, 0.01882$ & $(0.5622, 0.04928)$ \\
    \hline
    \ & $0$ & $(0.2148, 0.1249)$ \\
    $\nu = 3$ & $- \, 0.01$ & $(0.9738, 0.03968), \ (1.124, 0.01188)$
    \\
    \ & $[- \, 0.06, - \, 0.02]$ & no intersection \\
    \ & $- \, 0.0631$ & $(0.7409, 0.006101)$ \\
    \hline
    \ & $0$ & $(0.03893, 0.1222)$ \\
    $\nu = 4.5$ & $- \, 0.01$ & $(0.8903, 0.06232)$ \\
    \ & $- \, 0.02$ & $(1.123, 0.02343)$ \\
    \ & $[- \, 0.0858, - \, 0.03]$ & no intersection \\
    \hline
  \end{tabular}
  \caption{Intersection points from the Wilson loop phase transition
    curve (crossover) to the background phase transition curve (1-st
    order phase transition) in $(\mu; T)$-coordinates depending on
    the primary anisotropy $\nu$ and the magnetic field's anisotropy $c_B$.}
  \label{Tab:BB-WL}
\end{table}



\section{Electrical conductivity}\label{conduct}

Electrical conductivity contains information about transport and
optical properties of a physical system. Understanding the behavior of
the conductivity under different conditions is an essential step in
studying properties of the direct photons. 

\subsection{Conductivity in anisotropic background}

We use the model derived above to investigate properties of the direct
photons formed in HIC. The detailed consideration of the holographic
approach to electrical conductivity and direct photon emission rate of
QGP is presented in \cite{Iatrakis:2016ugz, Iqbal:2008by,
  Arefeva:2021jpa}. We briefly describe the framework for the
holographic investigations of the electrical conductivity below. \\

In the anisotropic background \footnote{Note that we get
  \eqref{eq:2.04} from \eqref{metric-full-anis} after $x_1\to x$,
  $x_2\to y_1$ and  $x_3\to y_2$. We call $x_1$ the longitudinal
  direction, i.e. the direction along the collision axis, $x_2$ and
  $x_3$ the transversal directions, and $(x_1 x_2 )$ are coordinates
  in a collision plane. The external magnetic field is aligned along
  $x_3$.}
\begin{gather}\label{metric-full-anis}
    ds^2 = \cfrac{L^2 \, \fb(z)}{z^2} \left[ - \, g(z)dt^2 
      + \mathfrak{g}_1(z)dx_1^2 + \mathfrak{g}_2(z)dx_2^2 
      + \mathfrak{g}_3(z)dx_3^2 + \cfrac{dz^2}{g(z)} \right]
\end{gather}
we consider a probe Maxwell field 
\begin{gather}
  S_{pert} = - \, \cfrac{1}{4}\int d^5x \sqrt{-g} f_0(\phi)
  F_{MN}F^{MN}. \label{pert}
\end{gather}
We assume this creates no gravitational backreaction.  Function
$f_0(\phi)$ is called a gauge kinetic function and is used to fit the
model with lattice data for the isotropic QGP at zero chemical potential
and zero magnetic field \cite{Iatrakis:2016ugz}. \\

Transport coefficients may be found within the linear response theory
using the Kubo relations. Namely, the electrical conductivity is
related to the retarded Green's functions of electric
currents. Therefore, one is interested in a plane wave solution to
\eqref{pert}, which will be holographically dual to $U(1)$ electric
current $J^\mu$ of the boundary theory. We take the ansatz of a plane
wave propagating along the $x_3$-direction. The wave amplitude is a function
of the ``holographic coordinate'' $z$ only. The Green's function in a
low frequency limit may be found either using the prescription by
Iqbal and Liu \cite{Iqbal:2008by} or Son and Starinets
\cite{Son:2002sd}. The equivalence of these approaches is demonstrated
in \cite{Arefeva:2021jpa}. The components of electric conductivity
tensor are then
\begin{gather}
  \sigma^{11} = \cfrac{2f_0(z_h)}{z_h} \,
  \sqrt{\cfrac{b(z_h)\mathfrak{g}_3(z_h)
      \mathfrak{g}_2(z_h)}{\mathfrak{g}_1(z_h)}}, \label{sigma11} \\
   \sigma^{22} = \cfrac{2f_0(z_h)}{z_h} \,
   \sqrt{\cfrac{b(z_h)\mathfrak{g}_3(z_h)
       \mathfrak{g}_1(z_h)}{\mathfrak{g}_2(z_h)}}, \label{sigma22} \\ 
    \sigma^{33} = \cfrac{2 f_0(z_h)}{z_h} \,
    \sqrt{\cfrac{b(z_h) \mathfrak{g}_1(z_h)
        \mathfrak{g}_2(z_h)}{\mathfrak{g}_3(z_h)}}, \label{sigma33}
  \end{gather}
where $z_h$ is the event horizon position \cite{Arefeva:2021jpa}.

\newpage

\subsection{Model tuning}

To tune the model parameters we use the results of lattice
calculations from \cite{Aarts:2014nba}. These results are known for the
isotropic QGP ($\nu = 1$) with a number of flavours $N_f=3$ at zero
chemical potential ($\mu = 0$) and without magnetic field ($c_B =
0$). Thus, we fit the behaviour of electrical conductivity to the
lattice data in this simplest case using the gauge kinetic function
$f_0(\phi)$. Then it is possible to calculate the conductivity in
anisotropic case and for non-zero parameters. We take the gauge kinetic
function in the following form 
\begin{gather}
    f_0(\phi) = \cfrac{0.0023}{\text{C}_{em}} \left[ 
      0.9 \exp \left(- \, \frac{\phi^2}{8.8} \right)
    + 0.11 \exp (- \, 0.05 \phi) \right],
\end{gather}
where $\text{C}_{em}=2e^2/3$ is electromagnetic constant for $N_f = 3$
and $\phi$ is the dilaton field. \\

\begin{figure}[h!]
  \centering
  \includegraphics[scale=0.55]{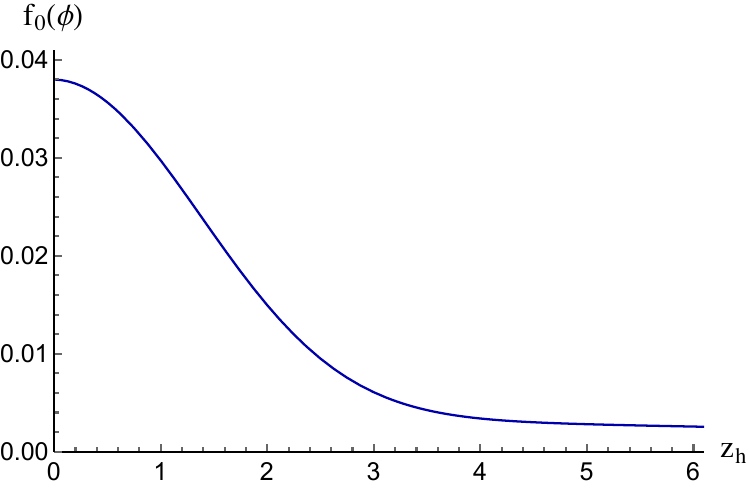} \quad
  \includegraphics[scale=0.55]{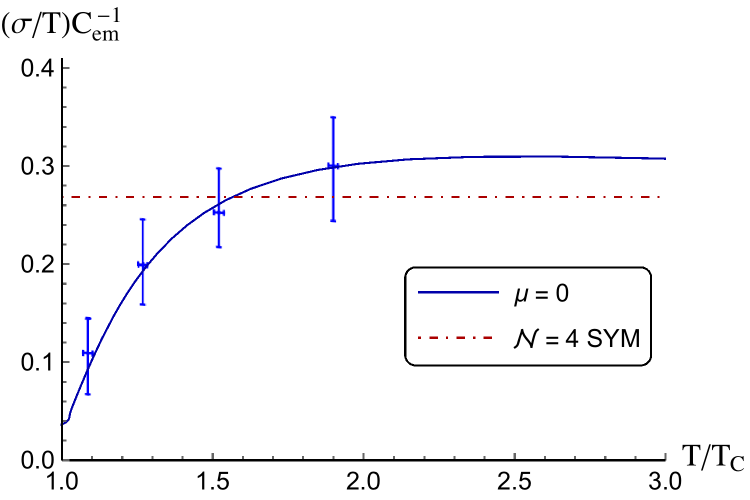} \\
  A \hspace{210pt} B \\
  \caption{Gauge kinetic function $f_0$ as a function of the horizon
    $z_h$ (A) and relation of electrical conductivity to temperature
    $\sigma/T$ as a function of the normalised temperature $T/T_c$ (B): numerical calculations in our model for $\nu = 1$, $c_B
    = 0$ and $\mu = 0$ (the blue curve), lattice calculations from \cite{Aarts:2014nba} (the blue
    dots with error bars) and the
    conductivity of $\mathcal{N} = 4$ SYM plasma (the red dash-dotted line).}
  \label{Fig:Tuning}
\end{figure}

Numerical calculation for the isotropic conductivity and the gauge
function are shown on Fig.\ref{Fig:Tuning}. Here the critical
temperature is $T_c = 0.155$ GeV. It was calculated as a temperature
of the BB phase transition for a given set of parameters. The conductivity
increases with temperature growth and saturates some value close to
the $\mathcal{N} = 4$ SYM plasma conductivity.

\newpage

\subsection{Isotropic case}

\begin{figure}[h!]
  \centering
  \includegraphics[scale=0.55]{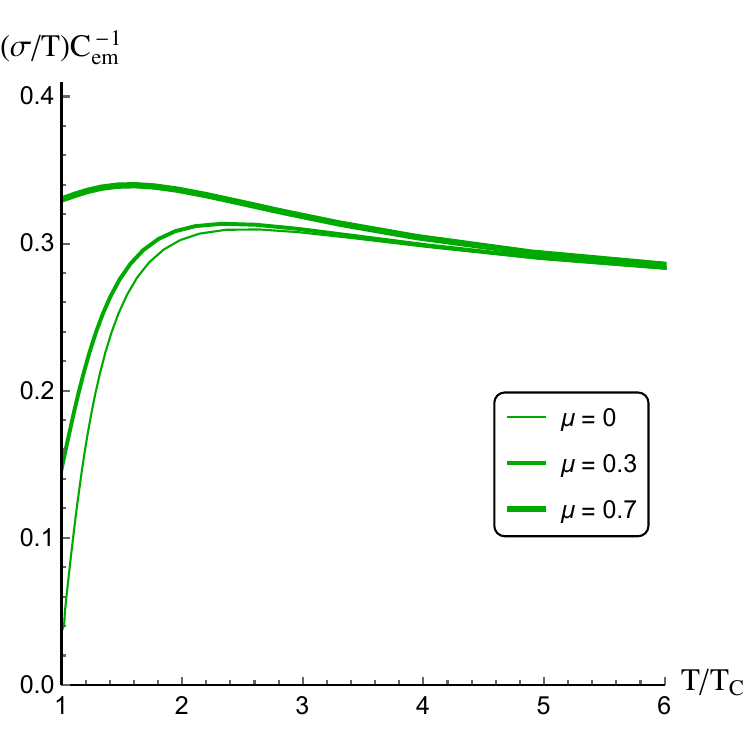} \quad
  \includegraphics[scale=0.55]{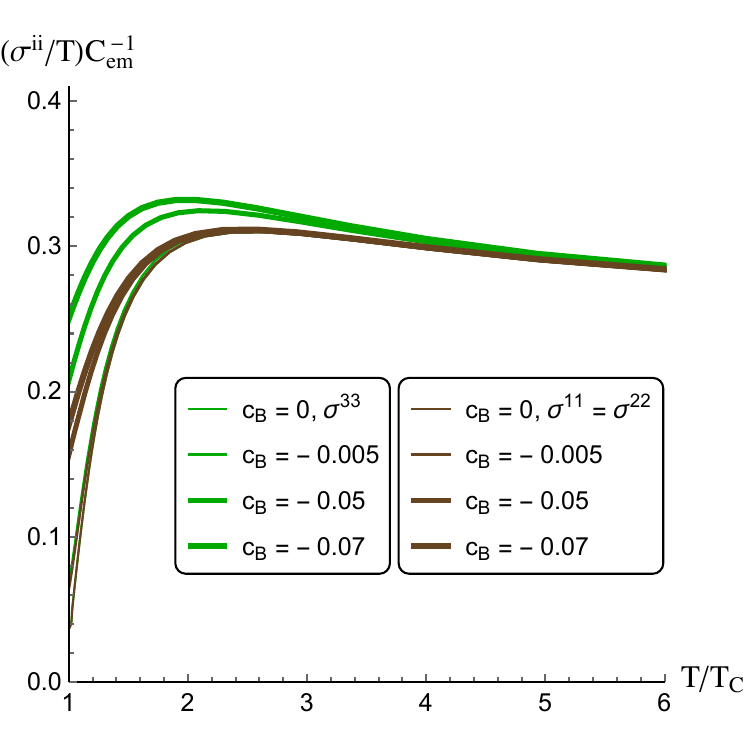} \\
  A \hspace{210pt} B \\
  \caption{The dependence of $\sigma/T$ on the
    normalised temperature $T/T_c$ in the isotropic case
    $\nu=1$ for zero magnetic field $c_B = 0$ with different values of
    chemical potential $\mu = 0, \, 0.3, \, 0.7$ (A) and for zero chemical
    potential $\mu=0$ with different
    values of the magnetic field parameter $c_B = 0, \, - \, 0.005, \, - \, 0.05, \, - \, 0.07$ (magnetic field is directed along the $x^3$): $\sigma^{33}$ (green
    curves) and $\sigma^{11}=\sigma^{22}$ (brown curves)~(B).}
  \label{Fig:n1b0mu}
\end{figure}

The conductivity tensor in the isotropic ($\nu = 1$) case with zero
magnetic field reduces to the single value $\sigma = \sigma^{ii}$
($i = 1,2,3$), so the QGP provides equal resistance to direct photons in all
directions. The dependence of the conductivity to the temperature ratio
$\sigma^{ii}/T$ on the temperature for different values of chemical
potential is shown in Fig.\ref{Fig:n1b0mu}A. The non-zero magnetic
field kills the degeneracy between two transverse  directions
(Fig.\ref{Fig:n1b0mu}B). The plots presented in Fig.\ref{Fig:n1b0mu}
demonstrate that the ratios $\sigma^{ii}/T$
grow with the normalized temperature till some temperatures around
$2T_c$. The precise location of the maximum depends on the chemical potential
and the magnetic field values. At high temperatures the $\sigma/T$ ratio decreases to some
constant values near the conductivity of $\mathcal{N} = 4$ SYM
plasma.  Plots on Fig.\ref{Fig:n1b0mu}A and \ref{Fig:n1b0mu}B
demonstrate the growth of conductivity with the magnetic field
and/or chemical potential growing for the same temperature values. Also, one
can see that in the directions orthogonal to the magnetic field the conductivity
is lower than in a parallel one (brown curves are lower than the
green ones of the same thickness). Therefore, QGP is more opaque along
the external magnetic fields.

\newpage

\subsection{Anisotropic case}

\begin{figure}[b!]
  \centering
  \includegraphics[scale=0.55]{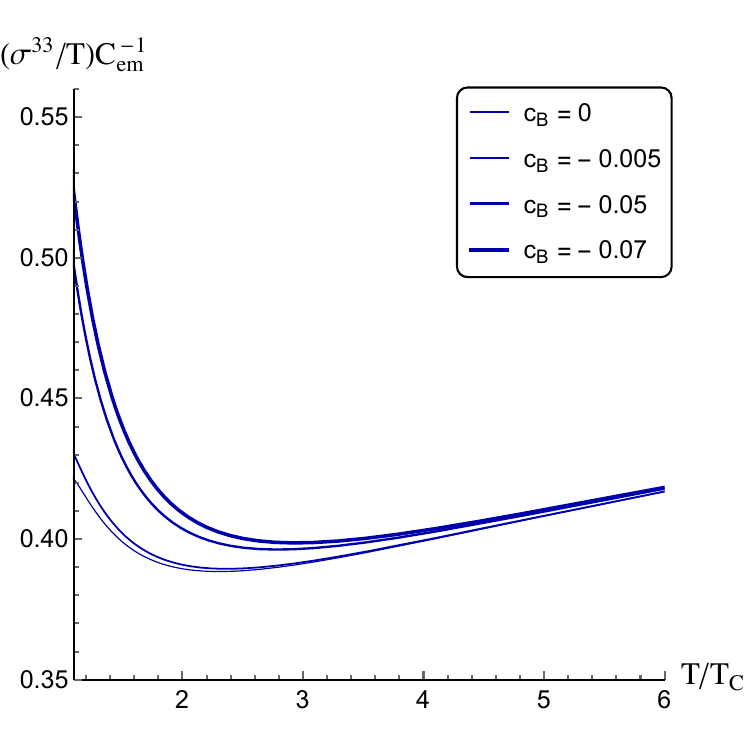} \qquad
  \includegraphics[scale=0.55]{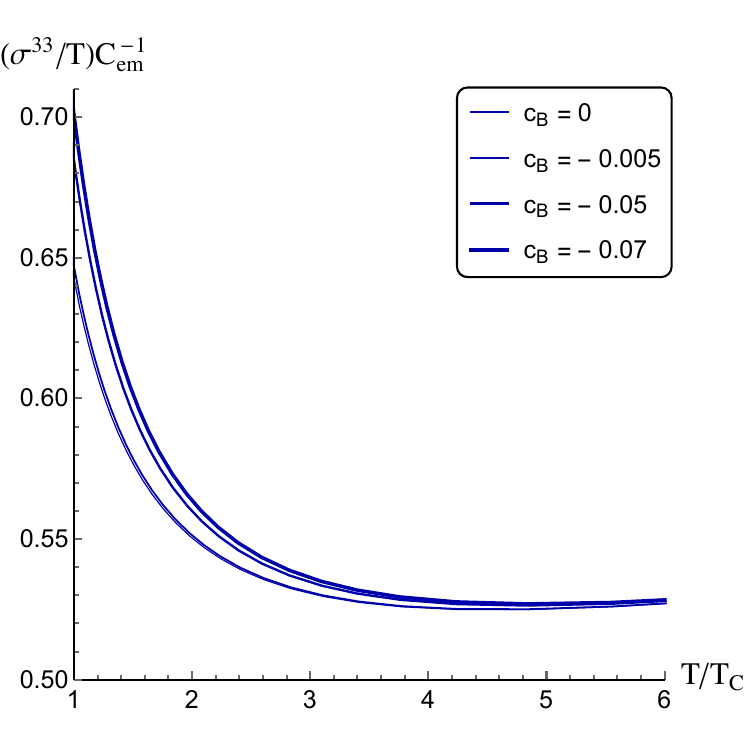} \\
  A \hspace{220pt} B \\
  \caption{The dependence of $\sigma^{33}/T$ on the normalised temperature
    $T/T_c$ in anisotropic case $\nu=4.5$ for the magnetic field parameter $c_B = 0, \, - \, 0.005, \, - \, 0.05, \, - \, 0.07$ with zero chemical potential $\mu=0$
    (A) and large chemical potential $\mu=1$ (B);
    critical temperature $T_c=0.121$ GeV.}
  \label{Fig:s33}
\end{figure}

In this subsection we study the electrical conductivity of QGP for
the anisotropy parameter value $\nu = 4.5$. The results of numerical
calculations of $\sigma^{33}/T$ (conductivity along the external magnetic
field) for different values of magnetic field and chemical potential
are presented in Fig.\ref{Fig:s33}. \\

We see that at the ratio $\sigma^{33}/T$ decreases with
temperature starting from $T = T_c$. The minimum position depends on the chemical
potential and the external magnetic field value. The $\sigma^{33}/T$  ratio after the minimum is greater for zero chemical potential
and for lower magnetic fields. Stronger magnetic field and larger
chemical potential increase the
conductivity at the fixed temperature. One can see that at high temperatures the curves in
Fig.\ref{Fig:s33} are getting closer to each other and almost
coincide. Thus, anisotropic QGP with $\nu =4.5$ along the external magnetic field direction is the most opaque at critical
temperatures, becomes more transparent around $T \approx 2 T_c$ for
$\mu=0$, Fig.\ref{Fig:s33}A, and $T \approx 3.5T_c$ for $\mu=1$,
Fig.\ref{Fig:s33}B, and after these values becomes more
opaque with increasing temperature again.


\begin{figure}[h!]
  \centering
  \includegraphics[scale=0.55]{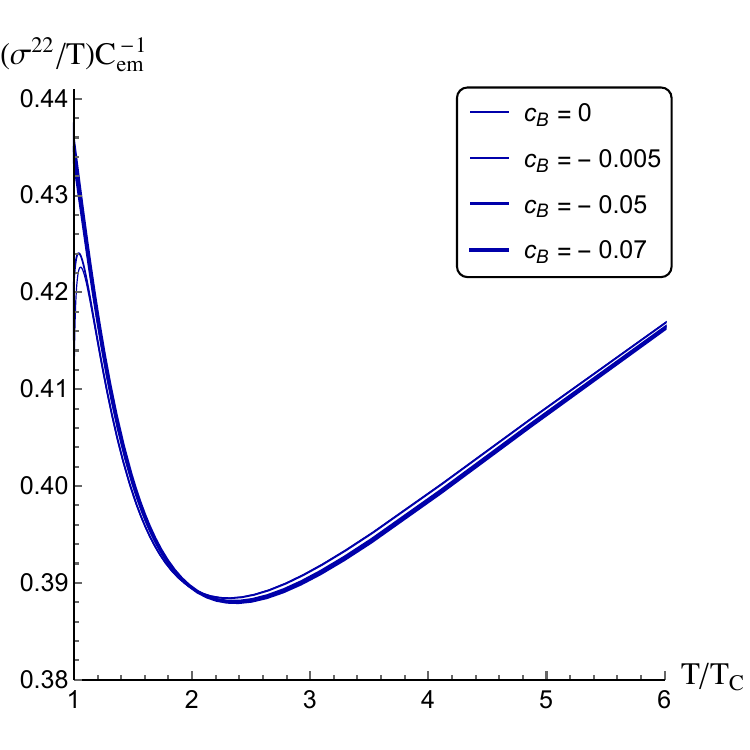} \qquad
  \includegraphics[scale=0.55]{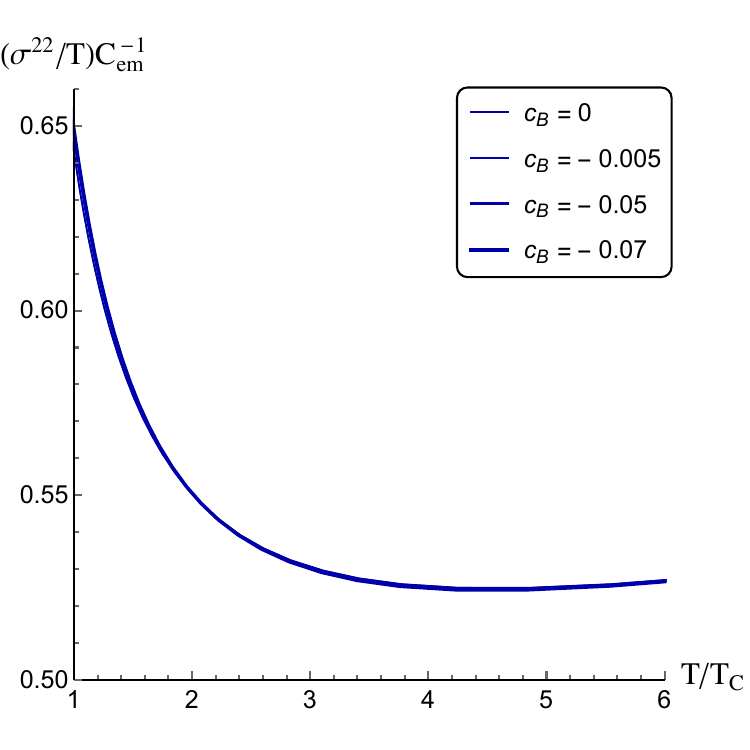} \\
  A \hspace{220pt} B \\
  \caption{The dependence of $\sigma^{22}/T$ on the normalised temperature
    $T/T_c$ in anisotropic case $\nu=4.5$ for the magnetic field parameter $c_B = 0, \, - \, 0.005, \, - \, 0.05, \, - \, 0.07$ with zero chemical potential $\mu = 0$
    (A) and large chemical potential $\mu = 1$ (B);
    critical temperature $T_c = 0.121$ GeV.}
  \label{Fig:s22}
\end{figure}
\begin{figure}[h!]
  \centering
  \includegraphics[scale=0.55]{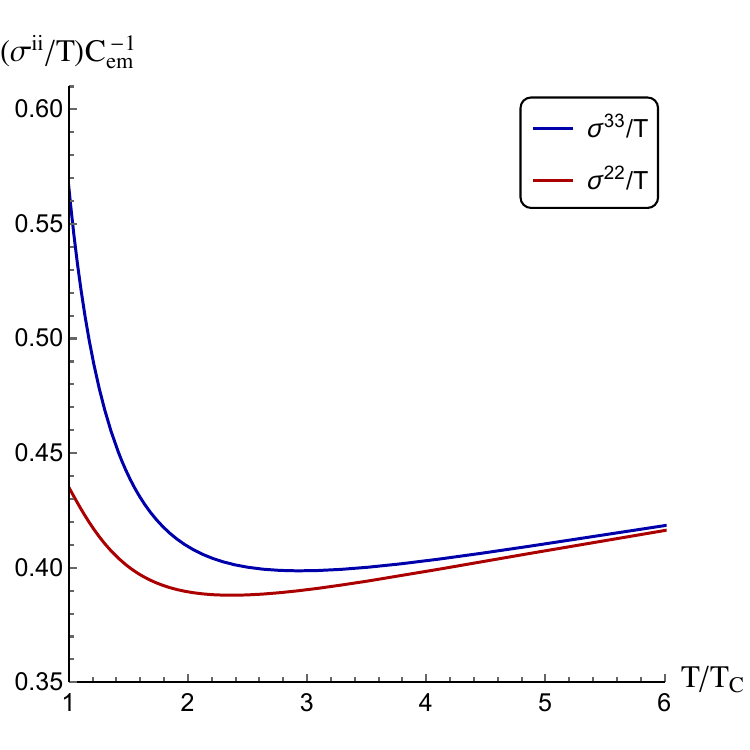} \quad
  \includegraphics[scale=0.55]{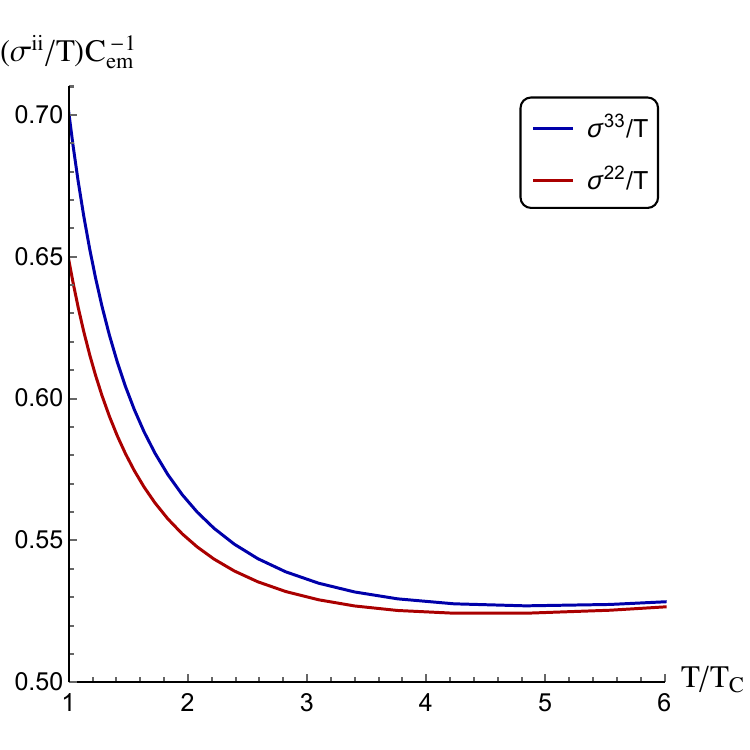} \\
  A \hspace{210pt} B \\
  \caption{The dependence of $\sigma^{22}/T$ (red curves) and
    $\sigma^{33}/T$ (blue curves) on the normalised temperature $T/T_c$ in anisotropic case $\nu=4.5$ for the magnetic field parameter $c_B = - \, 0.07$ with zero chemical potential $\mu=0$
    (A) and large chemical potential $\mu=1$ (B);
    critical temperature $T_c=0.121$ GeV.} 
  \label{Fig:s22s33}
\end{figure}

Thermodynamic properties of $\sigma^{22}/T$ are similar to those of
$\sigma^{33}/T$ and are shown in Fig.\ref{Fig:s22}. We see that near
the phase transition point $\sigma^{22}/T$ decreases to a minimal
value, and then starts to increase linearly at high
temperatures. Plots in Fig.\ref{Fig:s22}A,B show that higher the
chemical potential is, higher the ratio $\sigma^{22}/T$ becomes. However, the
growth rate at high temperatures is larger for $\mu=0$ case. The
curves for strong and weak magnetic fields are almost indistinguishable,
but the conductivity in stronger magnetic fields is greater than in weak ones. For zero magnetic field there is no difference
between $\sigma^{22}$ and $\sigma^{33}$, it can be seen in non-zero
magnetic fields only. The value of $\sigma^{33}/T$ is always larger than
$\sigma^{22}/T$. The comparison of these two conductivity components can be found on Fig.\ref{Fig:s22s33}. \\

\begin{figure}[t!]
  \centering
  \includegraphics[scale=0.55]{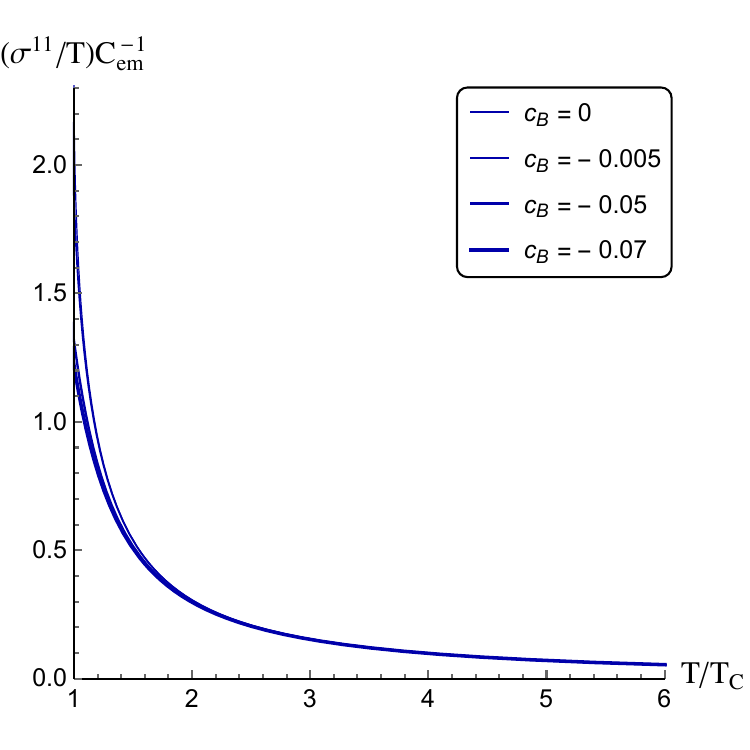} \qquad
  \includegraphics[scale=0.55]{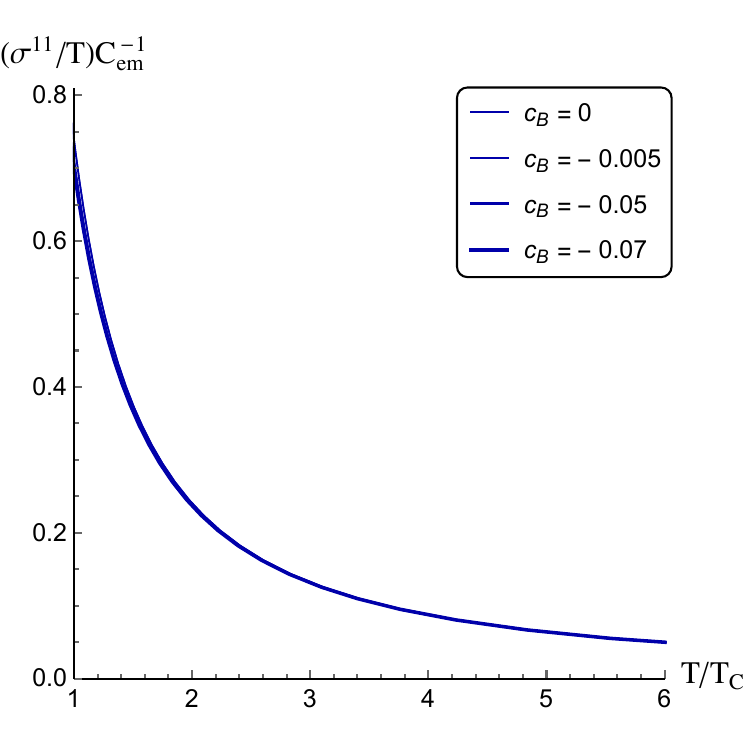} \\
  A \hspace{220pt} B \\
  \caption{The dependence of $\sigma^{11}/T$ on the normalised temperature
    $T/T_c$ in anisotropic case $\nu=4.5$ for the magnetic field parameter $c_B = 0, \, - \, 0.005, \, - \, 0.05, \, - \, 0.07$ with zero chemical potential $\mu=0$
    (A) and large chemical potential $\mu=1$ (B);
    critical temperature $T_c=0.121$ GeV.}
  \label{Fig:s11}
\end{figure}

The results of similar calculations for $\sigma^{11}/T$ are shown in
Fig.\ref{Fig:s11}. We see that  $\sigma^{11}/T$ monotonically
decreases to zero with temperature growth, while magnetic field and
chemical potential push it down. Or, in other words, QGP along the
collision line near the critical temperature is almost opaque and becomes transparent at high
temperatures. \\

\begin{figure}[h!]
  \centering
  \includegraphics[scale=0.4]{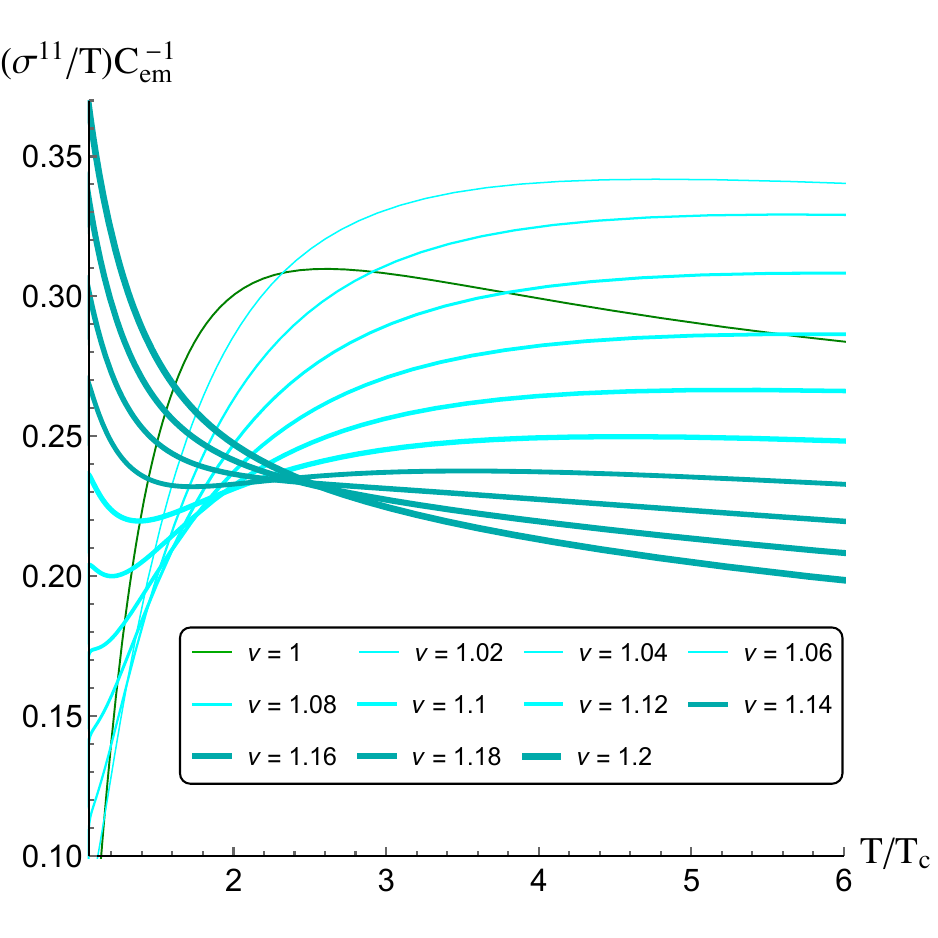} \qquad
  \includegraphics[scale=0.4]{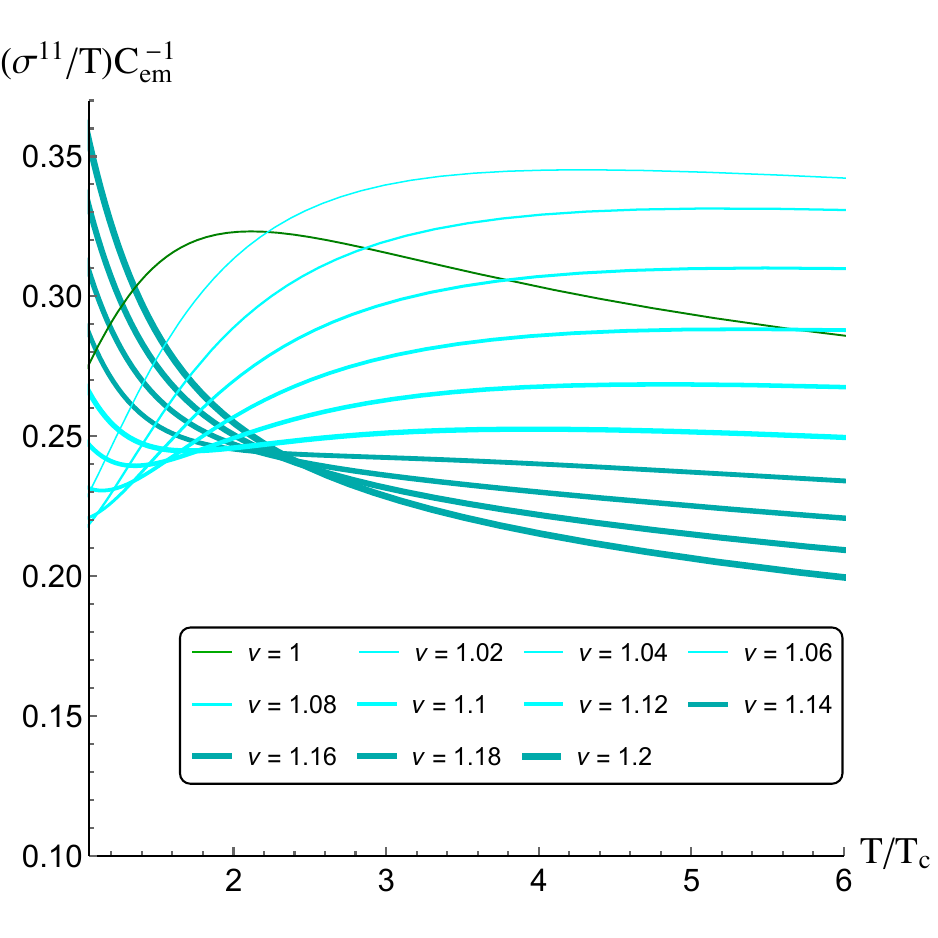} \\
  \includegraphics[scale=0.4]{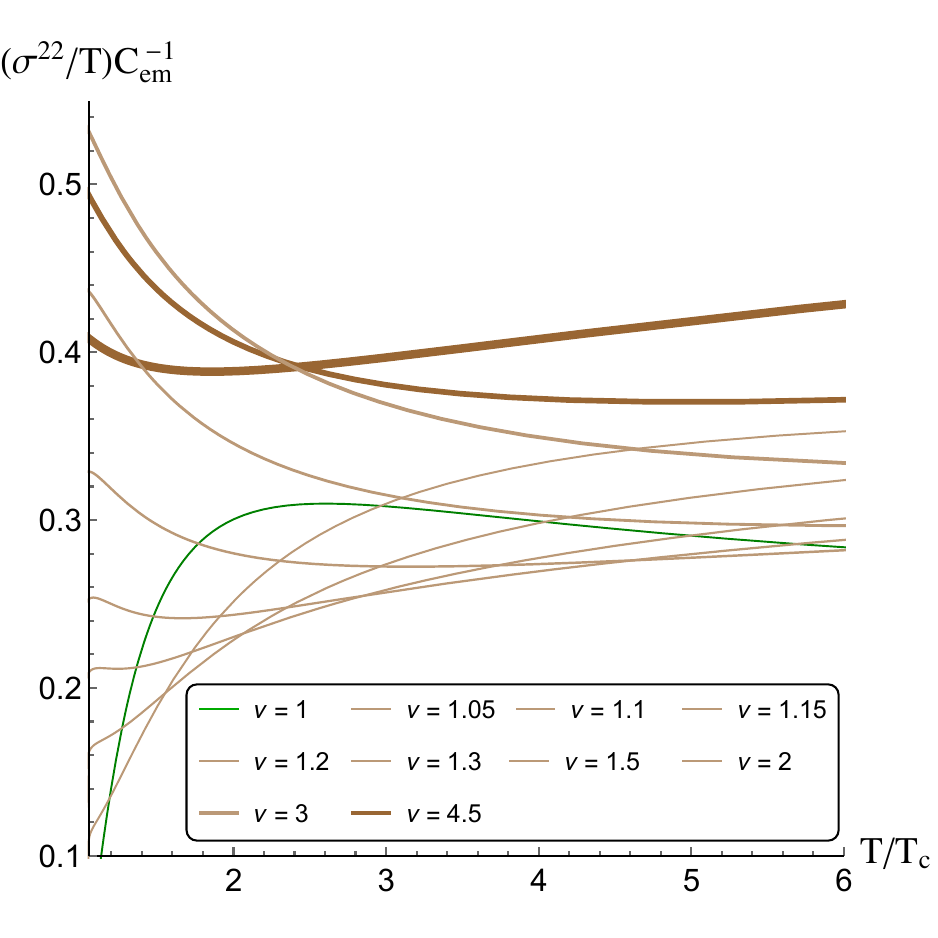} \qquad
  \includegraphics[scale=0.4]{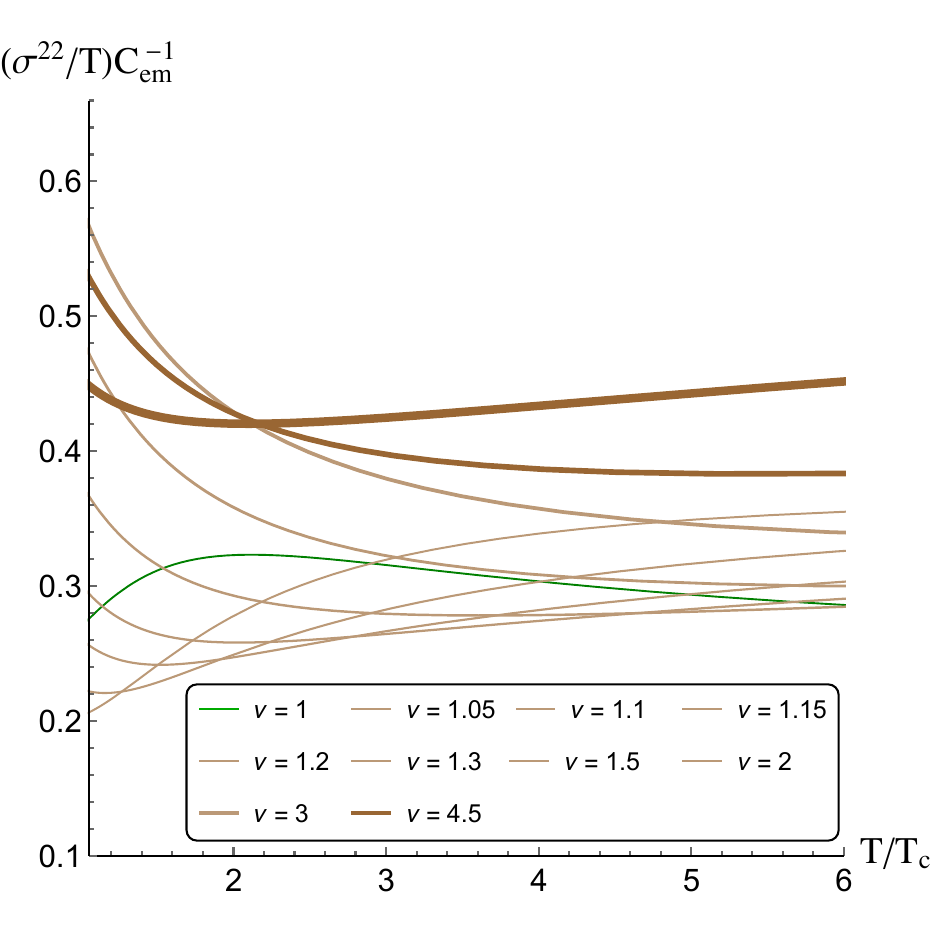} \\
  \includegraphics[scale=0.4]{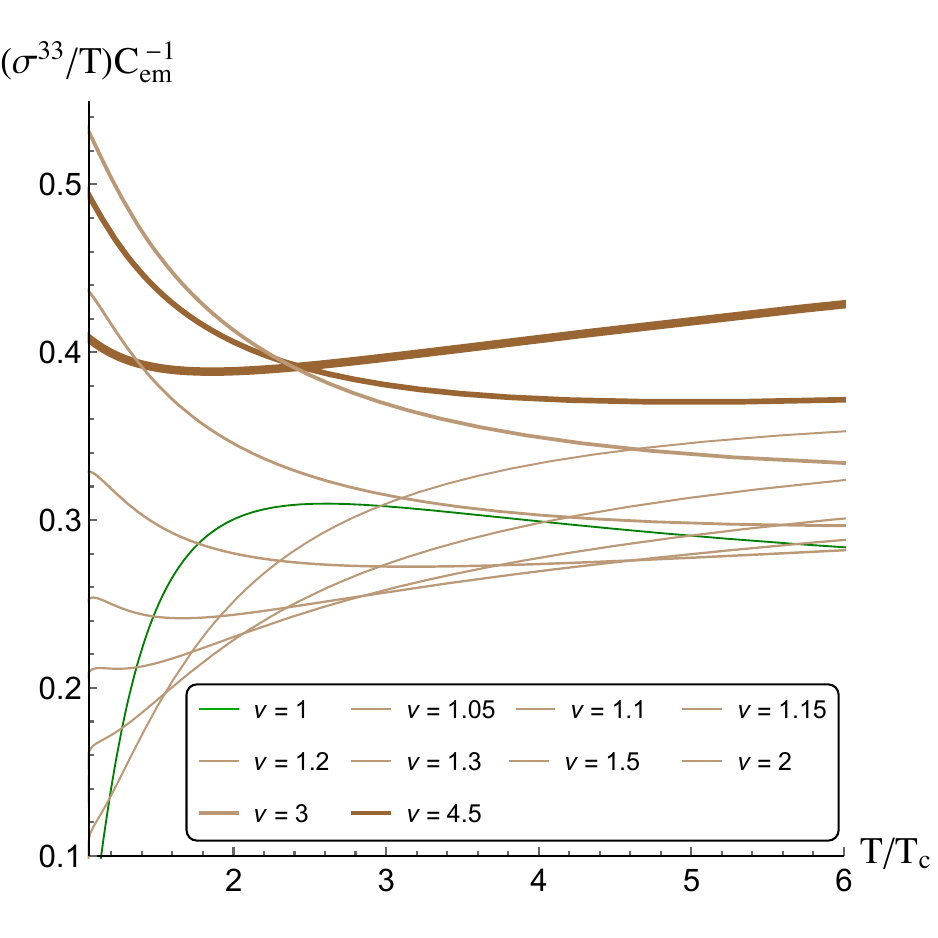} \qquad
  \includegraphics[scale=0.4]{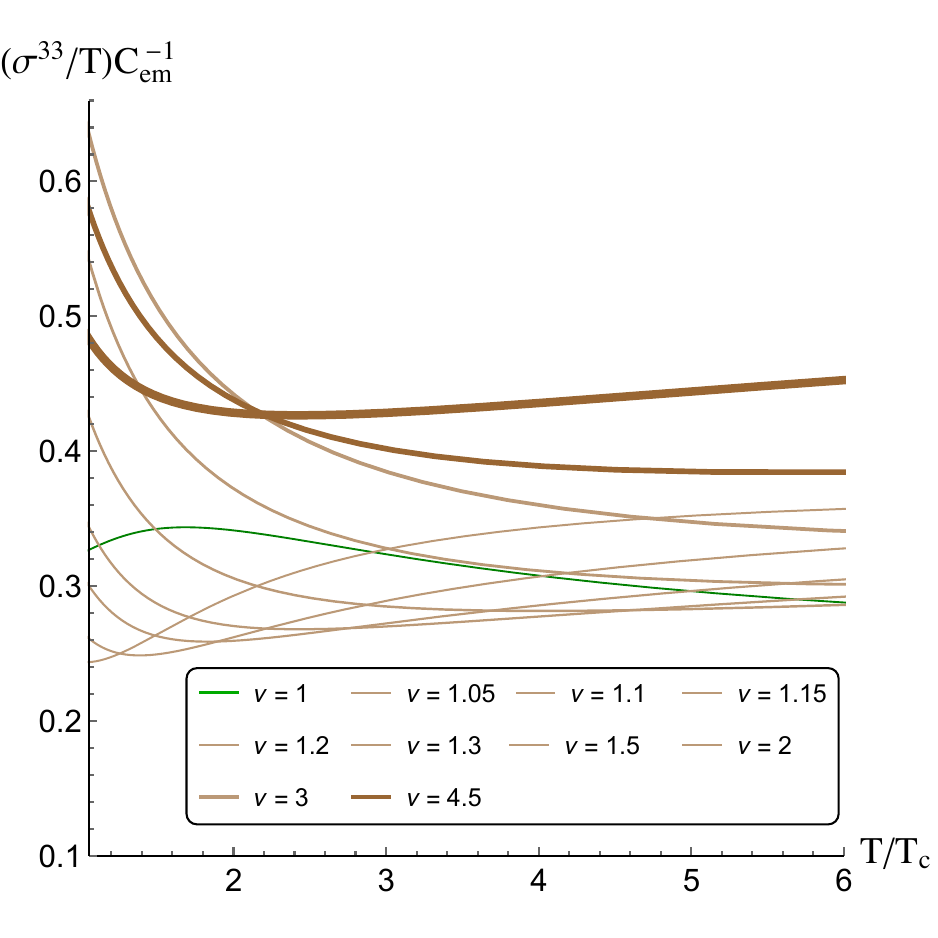}\\
  A \hspace{200pt} B
  \caption{The dependence of $\sigma^{11}/T$ (1-st line), $\sigma^{22}/T$ (2-nd line) and $\sigma^{33}/T$ (3-rd line) on the normalised temperature $T/T_c$ for different $\nu$ with $\mu = 0$, $c_B = 0$ (A) and $\mu = 0.5$, $c_B = - \,0.05$ (B).}
  \label{Fig:triptych23}
\end{figure}

On Fig.\ref{Fig:triptych23} one can see the change of the conductivity behavior with the primary anisotropy parameter increasing. The green curves denote the isotropic case $\nu = 1$ and for the anisotropic cases the thicker line corresponds to the larger $\nu \ne 1$. As anisotropy increases, qualitative changes occur in all directions, where as usually two transverse and a longitudinal directions differ crucially. At low energies conductivity in longitudinal direction increases at low temperatures and starts decreasing after approximately $T = 2.5T_c$. At high values of anisotropy parameter $\nu$ longitudinal conductivity decreases monotonically. On the contrary, transverse components increase at high temperatures. \\

\newpage

\section{Conclusion} \label{Sect:Conc}
\label{ConclDiscuss}

In this paper we have considered a twice anisotropic holographic model
for the light quarks \cite{Arefeva:2022bhx}. It is characterized by the
Einstein-dilaton-three-Maxwell action and the 5-dimensional metric with
the warp factor that has been considered in isotropic case for the light
quarks \cite{Li:2017tdz}. Our model has two different types of
anisotropy mentioned above: the anisotropy related to the parameter
$\nu$ (responsible for reproduction of the experimental energy dependence of
multiplicity), and the anisotropy related to the external
magnetic field (described by the third Maxwell field).  The
twice anisotropic model for the light quarks is much more complicated than
the twice anisotropic model for heavy quarks \cite{Arefeva:2021mag,
  Arefeva:2020vae}, since the deformation factor and the kinetic gauge
function $f{_1}$ have more compound form. \\

We have investigated the dependence of the phase transition structure
on the anisotropy, characterized by parameter $\nu$ and magnetic
fields. The phase transition structure is set up by the mutual
arrangement of two types of the phase transition lines -- the 1-st order phase
transition, originating from the metric background, and the crossover,
originating from the temporal Wilson loop behavior. We have got the
following results concerning the phase transition structure.

\begin{itemize}
\item For $\nu = 1$ the 1-st order phase transition in the light quarks
  model lasts from $(\mu_{CEP}, T_{CEP})$ to $(\mu_{max},0)$. Magnetic
  field $c_B \ne 0$ shortens the 1-st order phase transition,
  increasing $\mu_{CEP}$ and decreasing $\mu_{max}$. In the heavy quarks
  model \cite{Arefeva:2020vae} the 1-st order phase transition lied from
  $(0,T_{max})$ to $(\mu_{CEP\, HQ}, T_{CEP\, HQ})$ almost
  horizontally. Magnetic field shortened this line just shifting
  $\mu_{CEP\, HQ}$ to zero. For the light quarks model the 1-st order phase
  transition line disappears at $c_B$ about 7 times larger than it was
  for the heavy quarks model. Inverse magnetic catalysis is found for both
  models.
\item For $\nu > 1$ primary anisotropy decreases temperature of the 1-st order phase
  transition and increases $\mu_{max}$. It almost doesn't influence the
  $c_B$ value that the 1-st order phase transition line disappears at,
  like it was for the heavy quarks model \cite{Arefeva:2020vae}. 
\item For $\nu = 1$ magnetic field decreases the Wilson loop curve
  inclination until it becomes horizontal. This leads to the inverse
  magnetic catalysis effect for $\mu$ near zero and the direct magnetic
  catalysis for larger chemical potential values. In the heavy quarks
  model \cite{Arefeva:2020vae} the Wilson loop curve underwent the inverse
  magnetic catalysis only and shrinked tending to vertical $T$-axis. 
\item Primary anisotropy $\nu > 1$ decreases the Wilson loop phase transition temperature
  and the region of the direct magnetic catalysis. It also increases the absolute value of $c_B$
  at which the Wilson loop survives. For $\nu = 4.5$ the Wilson
  loop curve shrinks tending to horizontal $\mu$-axis. For the heavy
  quarks model the larger anisotropy was the weaker it decreased the
  Wilson loop temperature.
\item For $\nu = 1$ in the light quarks model the confinement/deconfinement
  phase transition is determined by the crossover from $\mu = 0$ to the
  intersection with the 1-st order phase transition $\mu_{by_2}$,
  where it picks up the main role. On the opposite, for the heavy quarks
  model \cite{Arefeva:2020vae} the 1-st order phase transition was
  decisive from $\mu = 0$ to the intersection with the crossover
  $\mu_{by_2}$.
\item Magnetic field $c_B \ne 0$ for $\nu = 1$ shifts the intersection
  point $(\mu_{by_2},T_{by_2})$ to the larger chemical potentials and
  the lower temperatures, but doesn't actually change the general scheme for the
  light quarks model. For the heavy quarks model \cite{Arefeva:2020vae}
  the 1-st order phase transition moved to above the Wilson loop curve and
  lost it's influence on the confinement/deconfinement process.
\item Anisotropy $\nu > 1$ makes the 1-st order phase transition  to move
  under the Wilson loop curve. Therefore the crossover determines the
  confinement/deconfine\-ment process from $\mu = 0$ to $\mu_{CEP}$, where
  the 1-st order phase transition begins and takes the main role with
  a jump. For the heavy quarks model \cite{Arefeva:2020vae} the 1-st order
  phase transition also moved under the Wilson loop curve and
  determined the confinement/deconfinement process from $\mu = 0$ to
  $\mu_{CEP\, HQ}$, where the 1-st order phase transition ends and
  gives the main contribution to the crossover with a jump.
\end{itemize}

We have also studied the DC conductivity of the QGP in the deconfined
phase under different conditions. The conductivity has different
values and thermodynamic properties in the HIC direction and in the
orthogonal plane presented, see Fig.\ref{Fig:n1b0mu}-\ref{Fig:s11}.

\begin{itemize}
\item For $\nu = 1$ 
  the conductivity
  \begin{itemize}
  \item $\sigma^{ii}/T$  
    near the critical temperature  grows with the temperature up to
    $\approx 2 T_c$ and then starts to decrease to a constant value
    near the conductivity of $\mathcal{N} = 4$ SYM plasma along all the directions;
  \item along the orthogonal to the magnetic field directions is lower
    than in a parallel one;
  \item grows with the chemical potential and the external
    magnetic field values.
\end{itemize}

\item For $\nu = 4.5$ the conductivity
  \begin{itemize}
  \item $\sigma ^{33}/T$ along the magnetic field direction has a minimum and increases at high temperatures (Fig.\ref{Fig:s33});
  \item $\sigma ^{22}/T$ along the transversal direction
    also has a minimum followed by the increment at high
    temperatures (Fig.\ref{Fig:s22}); $\sigma^{33}$ and $\sigma^{22}$ can be distinguished
    at the non-zero magnetic fields, and $\sigma^{33}$ is always greater
    than $\sigma^{22}$ (Fig.\ref{Fig:s22s33});
  \item $\sigma ^{11}/T$ along the longitudinal direction monotonically
    decreases (Fig.\ref{Fig:s11}).
  \end{itemize}
\end{itemize}

It is instructive to compare the conductivity for the light and heavy quarks.
Let us remind that for the heavy quark model we had \cite{Arefeva:2021jpa}
the following picture.

\begin{itemize}
\item For $\nu = 1$
  \begin{itemize}
  \item at temperatures higher than the critical value the ratios
    $\sigma^{ii}/T$ increase to the constant values depending on $\mu$
    and $c_B$;
  \item increasing $\mu$ or magnetic field increases the values of
    these constant (saturation) values;
  \item the conductivity $\sigma^{22}$ and $\sigma^{33}$ are almost
    indistinguishable for small values of the external magnetic field.
\end{itemize}

\item For $\nu = 4.5$ the conductivity
  \begin{itemize}
  \item $\sigma^{22}/T$ and $\sigma^{33}/T$ along the transversal directions
    increase with the temperature growth and saturate a certain value
    $\sigma^{22}/T\approx\sigma^{33}/T$ depending on the
    $\mu$ and magnetic field values;
  \item $\sigma^{11}/T$ along the longitudinal direction approaches zero at high temperatures.
  \end{itemize}
\end{itemize}

Thus, we see that for $\nu = 1$ the conductivity behavior for the
heavy and light quarks is different at near-critical temperatures,
but at high temperatures they saturate some constant values and these
values increase with the chemical potential and magnetic
field growth. The saturation constants are equivalent at zero $\mu$ and
$c_B$. Isotropic QGP for the light quarks as well as for the heavy quarks is
almost transparent in all directions near the critical temperature and
become opaque at high temperature. The main difference between
them may be seen at intermediate temperatures. \\

For $\nu = 4.5$ the conductivity behavior for the heavy and light
quarks is different at near-critical temperature as well as for
high temperatures. In particular, there is no saturation of
$\sigma^{33}$ and $\sigma^{22}$ at high temperature for the light quarks,
meanwhile these components of the conductivity for the heavy quarks reach a
constant value at high temperature. The longitudinal component of the
conductivity for the heavy quarks decreases to zero as it does for the light
quarks. \\

There are some obvious open questions regarding our model that are
worth exploring. Let us mention some of them. It would be interesting to study the quark potential for the light
quark model considered here like it was done for the heavy quark model
\cite{Arefeva:2019yzy}. \\

It would also be interesting to study the entanglement
entropy (EE) behavior for our model. Note that EE was studied in the
\cite{Arefeva:2018hyo} model and was not studied either in
the presence of an external magnetic field for the heavy quark model
\cite{Arefeva:2020vae}, or for the light quark model in the absence of an external magnetic
field \cite{Arefeva:2020byn}. \\

In the spirit of the jet quenching studying
\cite{Ageev:2017qpa, Ageev:2016gtl}  for the simplest
\cite{Arefeva:2018hyo} model, it would be interesting to investigate the
behavior of the jet quenching for the model considered in this article. Of
particular interest is the consideration of the external
magnetic field influence on the results. \\

In this paper we determine the MC/IMC as the increase/decrease of the
first order phase transition temperature with increasing the magnetic field value. In lattice calculations the MC and IMC are determined
as increase/decrease of the chiral phase transition temperature with
increasing the value of magnetic field \cite{Fukushima:2012kc,
  Andersen:2014xxa, Mamo:2015dea, Miransky:2015ava}. Note that the chiral phase
transition coincides with the first order phase transition in the
holographic model for heavy quarks \cite{Li:2020hau}, see also
studies of the chiral condensate in holography in \cite{Li:2016smq,
  Fang:2019xbk, Ballon-Bayona:2020qpq, Colangelo:2020tpr,
  Bohra:2020qom} and references therein. The study of the quark condensate and the
chiral phase transition is certainly of interest for the structure of the
phase transitions for light quarks.

\section{Acknowledgments}

This work is supported by Russian Science Foundation grant
20-12-00200.


\end{document}